\documentclass[aps, pra, twocolumn, preprintnumbers, amsmath, amssymb, nofootinbib]{revtex4} 

\pdfoutput=1

\usepackage[pdftex]{graphicx}
\usepackage{mathtools}
\usepackage[section]{placeins}
\usepackage{color}
\usepackage [english]{babel}
\usepackage [autostyle, english = american]{csquotes}
\MakeOuterQuote{"}

\begin{document}

\title{Defender Policy Evaluation and Resource Allocation Using MITRE ATT\&CK Evaluations Data\footnote{This work has been submitted to the IEEE for possible publication. Copyright may be transferred without notice, after which this version may no longer be accessible.}}

\author{Alexander V. Outkin$^{1}$}\email{avoutki [at] sandia.gov}
\author{Patricia V. Schulz$^{2}$}\email{tricia.schulz [at] outlook.com}
\author{Timothy Schulz$^{2}$}\email{tim [at] scythe.io}
\author{Thomas D. Tarman$^{1}$}\email{tdtarma [at] sandia.gov}
\author{Ali Pinar$^{1}$}\email{apinar [at] sandia.gov}

\affiliation{$^{1}$Sandia National Laboratories\\
$^{2}$Research performed while at Sandia National Laboratories}

\begin{abstract}
  Protecting against multi-step attacks of uncertain duration and timing forces defenders into an indefinite, always ongoing, resource-intensive response. To effectively allocate resources, a defender must be able to analyze multi-step attacks under assumption of constantly allocating resources against an uncertain stream of potentially undetected attacks. To achieve this goal, we present a novel methodology that applies a game-theoretic approach to the attack, attacker, and defender data derived from  MITRE\'~s ATT\&CK\textsuperscript{\textregistered} Framework.  Time to complete attack steps is drawn from a probability distribution determined by attacker and defender strategies and capabilities. This constraints attack success parameters and enables comparing different defender resource allocation strategies. By approximating attacker-defender games as Markov processes, we represent the attacker-defender interaction, estimate the attack success parameters, determine the effects of attacker and defender strategies, and maximize opportunities for defender strategy improvements against an uncertain stream of attacks. This novel representation and analysis of multi-step attacks enables defender policy optimization and resource allocation, which we illustrate using the data from MITRE\'~s APT3 ATT\&CK\textsuperscript{\textregistered} Evaluations.  
\bigskip
  

\end{abstract}

\maketitle

\section{Introduction}\label{sec:introduction}

Cyber defenders have to monitor and respond to attacks continuously, and their resource allocation decisions have to optimize against an uncertain stream of attacks. With no limits on possible start and end times of attacks, the attacker defender interaction becomes a game of indefinite duration where
the attack success probability or other metrics associated with a pre-defined time interval are not sufficient to measure the defender strategy effectiveness\footnote{Our approach relates both to dynamic game theory and Markov decision process. Those areas tend to use terms ``strategy'' and ``policy'' respectively to describe decision-making algorithms. We use terms policy and strategy interchangeably here.}.

The defender\footnote{We assume that both attackers and defenders can be individuals, groups, or organizations. This is why we sometimes refer to them as ``they''.} problem of securing the system becomes the problem of resource allocation across different parts of the system, where the defender policy objective is to allocate resources to detection and assessment and to place sensors across the defender system while keeping the expenditures  below a certain threshold. The top-level defender problem is then to design a system that provides adequate response and warning about attacks, sufficient automated prevention or response against identified attacks, and uses the available resources in an optimized fashion.

Real world attacks generally involve multiple steps. The attacker needs to get into the defender system, move through it, gather additional information, communicate with a command and control center, and act on objectives. This multi-step nature of attacks increases the scope of the defender problem, but also presents multiple detection and response opportunities at different steps in the attack. However, without quantitative metrics that reflect the effects of the defender resource allocation, the defender has no reliable criteria for allocating resources across such detection opportunities. Should the defender spend time educating system users against phishing attempts, investigating lateral movement events, looking for command and control messages, etc.?

We present an approach for solving this problem. It enables quantitative evaluation of the defender policy against multi-step attacks. Our approach calculates attack success metrics, such as time-to-success distribution and steady state distribution,
as a function of attacker and defender capabilities for each  attack step. It quantifies the effects of a defender allocating more resources for detection at specific attack steps by calculating the sensitivity of attack success metrics to such changes. This tradeoffs evaluation allows defenders to understand where they will get the best return from improving detection and therefore increasing attack difficulty. Furthermore, the defender can continually evaluate their allocation decisions against the latest threat information and improve their defense posture with future investments.

This approach allows incorporating real-world data, such as the results of MITRE ATT\&CK\textsuperscript{\textregistered} Evaluations\footnote{https://github.com/mitre-attack/attack-evals} (hereafter, ATT\&CK Evaluations) into attack success metrics and defender response evaluations. As importantly, it allows aggregating data from empirical observations associated with parts of an attack, such as ATT\&CK Evaluations or cyber experiments, into metrics describing an attack as a whole. We describe the theoretical model and demonstrate our approach using the APT3 ATT\&CK Evaluations data overlaid on an attack scenario developed on the basis of a simplified APT3 attack on a cyber-controlled power grid. In this scenario, an initial infection occurs in an enterprise network, pivots to the power grid control network and affects grid control devices in an attempt to disrupt grid operations. As in many real world scenarios, the grid operator has no control on when an attack can start and end or when the attacker who is already in the system and is ready to attack will execute on his objectives.

We show how streams of multi-step attacks can be analyzed by representing them as Markov processes by extending the GPLADD framework introduced in \cite{Outkin2019} to games of indefinite duration. We approximate GPLADD games by Markov chains with states being the success conditions in the attack graph. In certain situations, GPLADD games can be represented by Markov chains exactly, in other situations Markov chain approximation may be less precise than a closed-form or a Monte Carlo simulation. This can happen when the attack graph has a complicated structure or when agent strategies have long dependencies on attack history. In practice, non-Markovian dynamics can be converted to Markovian by expanding the state space. Markov processes are often used to approximate more general stochastic processes. For example \cite{Hobert2006} provides a method for generating approximations of intractable probability distributions using Markov chains. Yet, even approximate Markov chain representation is valuable because it allows for analyzing and visualizing the attacker-defender game progress quickly and explicitly, evaluating tradeoffs between detection expenditures at different parts of the defender system, improving sensor placement and assessment strategies, and developing an optimized defender strategy.

In this paper, we describe the following:
\begin{enumerate}
\item Attack success conditions representation using GPLADD.
\item Attack dynamics approximation using Markov chains.
\item Attack metrics evaluation
\item An example phishing-based attack chain represented using the APT3 ATT\&CK Evaluations data.
\item The methodology for inferring Markov chain transition rates from the ATT\&CK Evaluations data.
\item Sensitivity of attack success metrics to detection at different steps of the attack.
\item Analysis varying defender capabilities and their impact on attack success metrics and the transition diagrams.
\item Applications to defender policy optimization.
  
\end{enumerate}

A key part of our approach is the ability to evaluate the attack success metrics in regard to different defender strategies and resource allocations. We show quantitative evaluation of the overall attack success probability, time the attacker spends in the ``Ready'' (completed) state, and evaluate effects of different defender mitigations and strategies. The ``Ready'' state residence time is important, because it represents a practically important metric of how well the defender system is protected and because it reflects the ability of the attacker to coordinate multiple attacks if they have the ability to keep them in the ``Ready'' state with low probability of disruption by defender.

Our novel contributions include:
\begin{itemize}
\item A method for representing attacker defender contests as games of indefinite duration.
\item Introduction of the fraction of the time that attack can achieve in the completed or ready state (``Ready'' state residence time) as a metric to evaluate defender strategy in games of indefinite duration.
\item Explicit estimation of time-to-success distribution for the entire attack.
\item A method for representing ATT\&CK Evaluations data as GPLADD games and Markov chains.
\item A method for generalizing defender capabilities from ATT\&CK Evaluations detection categories.
\item A method for estimating the quantitative success parameters for attacks.
\item A method for evaluating the effects of defender investments in detection at different parts of the defender system by evaluating the attack success metrics and attack transition diagrams.
  
\end{itemize}

This paper is structured as the following: section \ref{sec:priorWork} describes relevant literature. Section \ref{sec:approach} describes our approach. We start with a detailed representation of attack as GPLADD games as described in \cite{Outkin2019} and then show how this representation can be transformed into a Markov chain representing the attack dynamics, where the transition probabilities are affected by the defender strategy. Section \ref{sec:AttackAndResults} shows the analysis of a multi-step attack, similar to APT3 as it is described in MITRE ATT\&CK framework and using the MITRE ATT\&CK Evaluations data. This example starts as an attack against an enterprise network with the goal of pivoting and acquiring control of an RTU (Remote Terminal Unit) on a SCADA/ICS system. The attacker attempts achieving his goals by initiating attacks against individual users of the enterprise network, in what we call an attack campaign composed from multiple single attacks. The Conclusions section summarizes and outlines possible extensions to this work.

\section{Prior Work} \label{sec:priorWork}

Our theoretical approach builds on and extends our work building attack models as GPLADD games (\cite{Outkin2019}). GPLADD is a game-theoretic approach to represent the attack success conditions as attack graphs and to quantify the attack success metrics as a function of the attacker and defender strategies in continuous time. In this paper, we extend and approximate GPLADD as discrete-time Markov chains as described in \ref{sec:approach}.

GPLADD is a dynamic game between attacker and defender. GPLADD is formally specified by the attack structure, and the attacker and defender strategies and actions at different stages of the attack. GPLADD treats time explicitly as a physical constraint and represents explicitly and  probabilistically the time different attack steps take. It and provides for a rich and detailed representation of attack states steps and the time they take, and detailed attacker and defender strategies, and attack analysis. It allows calculating the attack probability of success, probability of detection, costs, and other parameters quantifying the attack dynamics. We briefly describe GPLADD here, closely following \cite{Outkin2019}.

It was originally introduced as a finite duration game to represent trust in outcomes of development processes, where the defender has control .
We extend it in the Section \ref{sec:MarkovApproximation} to represent games of indefinite duration to account for difference between trust and security, and to include detection.

GPLADD can embody a specific attack, represent attack and system parameters and agent strategies, and therefore enable evaluation of the defender strategy and attack success metrics estimation.

The literature on Markov chains is vast. Markov chains applications to represent agent interactions include \cite{Banisch2015} and  \cite{Feng2017} as well as literature on Partially Observed Markov Decision Process (POMDP) (\cite{Miehling2018pomdp}). Paper \cite{Miehling2018pomdp} provides a model of an attacker-defender interaction in represented as a POMDP from the point of view of a defender. \cite{Miehling2018pomdp} has a somewhat similar formalism as in \cite{Outkin2019} to represent the attack success conditions, which the authors call a ``condition dependency graph''. It models the dependency between the attacker progress and the resulting system state and the set of the available exploits available from the attacker. A fundamental difference between the approaches in \cite{Miehling2018pomdp} and \cite{Outkin2019} is that the latter treats the time required to complete an attack step explicitly. The POMDP-based approaches explicitly treat the uncertainty in the defender knowledge in the system state and the resulting cyber awareness. A full review of the corresponding literature is outside the scope for this paper. The literature on cyber awareness include \cite{Wollaber2019} that describes the implementation of their approach on an high performance computer. \cite{Wagner2019} describes a defender strategy based on network segmentation. Game theory and advanced persistent threats (APTs) are discussed in \cite{Huang2020}.

The MITRE ATT\&CK Framework  is described in \cite{Strom2018}. It contains information on observed adversary tactics, techniques, and procedures (TTPs). MITRE leveraged the information in this framework to organize their ATT\&CK Evaluations, which tested the capabilities of EDR (Extended Detection and Response) vendors to detect TTPs associated with specific adversary campaigns. The results of these evaluation provide information that can be used to estimate attacker success and detection probabilities on different steps in an attack. APT datasets are also discussed in \cite{Stojanovic2020}. Concrete steps and attacker options at different stages of the attack can be informed from MITRE ATT\&CK (see \cite{Strom2018, Strom2017, MITRE2020} for more information). Additional empirical data on attackers and defenders might be generated using emulation (\cite{leeuwen2010}, see also \cite{Leeuwen2016, jones2017, dewaard2017, urias2016}).

The kill chains described in \cite{Hutchins2011} lay out  the logical phases in attack progress, as well as provide a high-level framework for reasoning about attacker and defender actions, attack progress, and detection. However, kill chain representation itself does not provide a quantitative framework for attack analysis or for specifying concrete steps in execution of a particular attack. Kill chains do not map directly into GPLADD; they represent a different, higher level of abstraction that may not represent the execution steps, strategies, timings, or dependencies.

Other papers that address detection and take into account the dynamic nature of the attacker defender interaction include \cite{Nguyen2008}, \cite{Zhu2008}, \cite{Alpcan2003} and \cite{Alpcan2004}. They deal with the problem of detection related tradeoffs and sensor data aggregation in a dynamic game setting, somewhat similar to GPLADD. Unlike GPLADD, they do not represent the steps in the attack.

The approaches in \cite{wilt2008} use detection and false alarm benchmarks to quantify confidence levels in system verification procedures. Hardware Trojan detection methods include \cite{agrawal2007}, \cite{jin2008}, and \cite{zhang2015}, and others. The \cite{pentrack2015} represent trust in the system as the product of individual subcomponents trust levels. Additional work on the attacker-defender interaction includes \cite{graf2016, graf2016b} and \cite{kamhoua2016} which optimize the Trojan detection technologies.

The security metrics literature includes \cite{Pendleton2016} and work attacks and attack trees modeling that include \cite{Schneier1999}, \cite{McDermott2001}, \cite{Moore2001}, \cite{Swiler2001}, \cite{Ingols2009}, \cite{noel2010}, \cite{Poolsappasit2012}. 

\section{Approach} \label{sec:approach}
We start with a GPLADD game and translate it into a Markov chain representation. GPLADD games were introduced in \cite{Outkin2019} and applied to Markov Games in \cite{Bishop2020}. A GPLADD game represents the attacker-defender game as a general stochastic process that is determined jointly by the attack graph and the attacker and defender capabilities and strategies. We introduce the Markov chain representation of attacker-defender interaction in section \ref{sec:MarkovApproximation}. We then develop a Markov chain representation of a simplified APT3 attack\footnote{We will call this a ``notional attack'' in what follows to indicate that we represent a simplified version of an APT3 attack.} in section \ref{sec:AttackAndResults} and develop a method to estimate the transition probabilities based on attacker and defender capabilities and available data. We then apply our methodology using results from the APT3 ATT\&CK Evaluations.

We assume that the defender constantly monitors for the attacker activity associated with some or all steps and execution methods of the attack using the defender IDS system and by collecting  sensor information to infer and observe the attacker activity. This detection is fine-grained in the sense that defender is able to associate detection events with possible attack steps. The defender may not have the capability to disrupt the attack completely, and so may  disrupt the attacker progress only partially, for example by identifying and cleaning the malware used for a particular lateral movement. Disrupting the attack completely sends the attacker to the beginning of attack, and disrupting the game partially sends the attacker to one of the previous steps of the game.

In a PLADD game, the defender's attack disruption is considered a ``take move'', that sends the attacker to one of the previous steps of the attack or to the beginning of the attack. We assume the defender can take immediate control of their system, incurring appropriate cost for that action (such a cost may include rendering the system inoperable for a period of time). The ability to execute effective and timely take moves is determined by the ability to detect the attacker activity, and associate it with an attack step and a part of the defender system, and may vary substantially across different defenders.  Detection is probabilistic and is affected by the defender capabilities and allocation of resources to sensing and assessment of specific attack indicators. Allocating more resources to detecting specific events may increase the probability of correctly detecting the attacker activity, but it also increases the defender costs and potentially reduces resources available to other parts of the system. 

GPLADD represents multi-step games, where the attack graph is represented as a set of necessary conditions for the attack success. The attacker strategy describes how the attacker will go about achieving these conditions. As in \cite{Outkin2019}, we focus on quantifying individual attack parameters to enable the attack ranking both from the point of view of the attacker and the defender. We demonstrate using Markov chain approximation of GPLADD attacks how the defender can allocate their resource to diminish attack effectiveness.

A GPLADD game consists of multiple PLADD games (defined in \cite{Jones2015}) connected by a graph structure and other elements.
A PLADD game is a contest for control of a single resource by the attacker and the defender. In PLADD, the defender has one or more different moves (take and morph in the original PLADD) to assert control over the resource. The attacker has only one move that represents the initiation of the attack. We also call it a ``take'' move. The attacker take move does not succeed immediately, but rather after a period of time termed time-to-success that is represented as a random variable with the time-to-success distribution denoted as $f(t)$.

This time-to-success distribution represents the difficulty of the corresponding stage in the attack. It reflects both the attacker capabilities and the defender system and security measures in place. An example determination of time-to-success distribution in application to a specific example is described in \cite{Chen2019}. A study by \cite{Holm2014} evaluates  possible distributions and their parameters for time to compromise of a computer system based on empirical data. 

Both the attacker and defender incur costs for executing their actions. These costs can be fixed or dependent on action duration or other factors.

\begin{figure}
 \centering
 \includegraphics[keepaspectratio=true, scale=0.25]{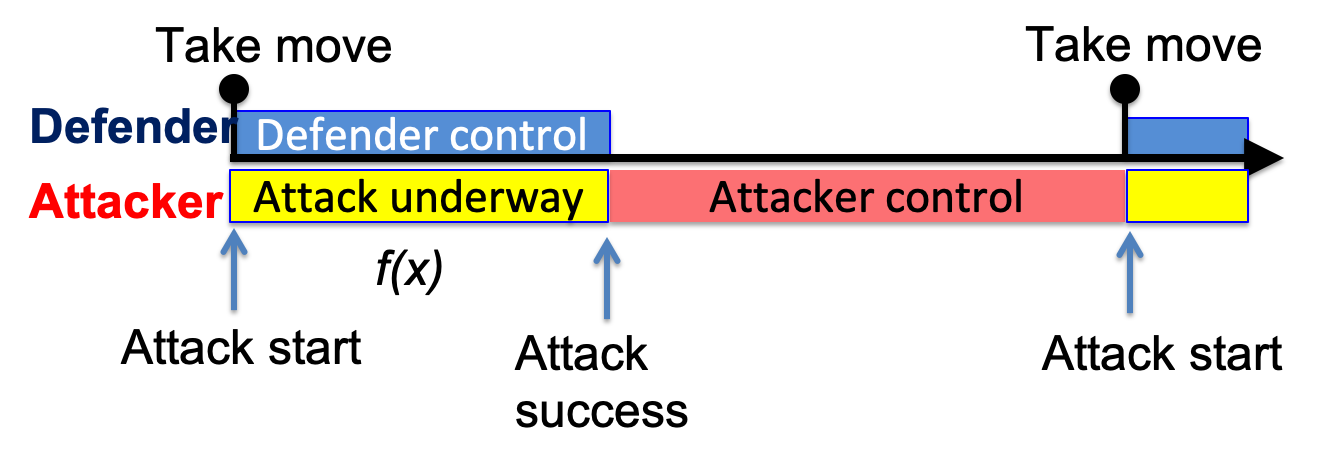}
 \caption{PLADD Game:  blue indicates defender control, red shows attacker control, and yellow shows attack in progress.}
 \label{fig:PLADD_Game}
\end{figure}

This attacker-defender contest can therefore be represented as a sequence of attacker and defender moves that may be executed without the full knowledge of the system state. For example, the defender may make the take move on a system that is not under attacker control. This interaction is illustrated in the in Figure \ref{fig:PLADD_Game}. It represents alternating periods of the defender control (blue bars above the x-axis) and attacker control (red bars below the x-axis), .

In GPLADD an attack is a partially ordered collection of PLADD games. These PLADD games can take place on certain parts of the defender system or outside of it. GPLADD distinguishes explicitly between an attack and attack graph. An attack is a game between attacker and defender that takes place on attack graph. The game dynamics is affected by the attacker and defender strategies. 

An attack $\Psi$ is defined as the following: $\Psi = \{\psi, \zeta^H, \zeta^D\}$, where $\psi$ is an attack graph,  and $\zeta^H(\psi)$ and $\zeta^D(\psi)$ are the attacker and defender strategies.

For the attack $\Psi$ to succeed, the attacker must achieve a set of conditions defined over a set of nodes $V_{\psi}$ that is composed of the nodes in $V$ in the defender system denoted as $v_\psi$ and  may involve the nodes outside of the defender system to represent the actions executed by the attacker elsewhere. The latter are denoted as $v'_\psi$ with $V_\psi = v_\psi \cup v'_\psi$. The attacker may complete certain parts of the attack outside of the defender system. For example, the attacker may modify a file using his own system after first stealing the file from the defender, with the goal of later injecting the modified file. An example attack and a more extensive discussion on the attack graph specification can be found in \cite{Outkin2019}.

The attack graph $\psi$ represents both the success conditions as the nodes in $\psi$ and the logical dependency between different steps in the attack as the links on the graph. These steps are represented as a set of directed acyclic chains. For attack to succeed, the attacker must complete all the steps.

The set of terminal nodes of all those chains in the attack graph constitute the necessary and sufficient conditions for the attack success.

A single chain $l$ in the attack graph $\psi$ is $\{c_1^l, \ldots, c_{n_l}^l\}$, where the superscript $l$ signifies that this is a $l$'th chain in the attack graph $\psi$. Each element in $l$ is a logical condition on $V_\psi$. The term $c_{n_l}^l$ denotes the terminal node in chain $l$. The numbering of the chains does not imply any ordering or temporal precedence.

The attack graph $\psi$ is represented as follows:

\begin{equation}
\psi= \{\{c_1^1, \ldots, c_{n_1}^1\}, \ldots, \{c_1^m, \ldots, c_{n_m}^m \} \},
\end{equation} 

A condition $c_i^j$ in $\psi$  may represent a specific system state or a change in the system state. For example, the logical condition for attacker control on node $v_i$ is represented as $s(v_i) = 1$ where $s$ is the state of the node and can take three values: $\{0, 1, 2\}$, where 0 denotes defender control, 1 stands for attacker control, and 2 for attack in progress.

Having all terminal conditions in $\psi$ being satisfied at the same time is necessary for the attack success. This is expressed as: 

\begin{equation} \label{eq:attSuccess}
S(\psi) = \{c_{n_1}^1 \land \ldots  \land c_{n_m}^m\},
\end{equation}
where $S$ represents ``attack success'' and $\land$ represents logical ``AND''. 

Once started, the attack on node $v_i$ succeeds after a certain delay, which we call the time-to-success. The time-to-success can be be either deterministic or probabilistic.  The time-to-success is a node-specific random variable distributed according to $f_i(\cdot) \text{ for all } v_i \in V_\psi$.

The time-to-success distribution reflects features of the node, of the defender system, as well as the defender and attacker capabilities.

The attack graph by itself does not fully determine the paths through inside or  outside of the defender system that the attack may take. These paths and the attack dynamics are jointly determined by the attack graph and the attacker and defender strategies.  The attack dynamics can have arbitrary complicated structure determined by the attack graph and the attacker and defender strategies. Unlike the attack graph, the attack pathways do not need to be acyclic. This enables generating the attack pathways as an emergent property of the attacker-defender interaction, rather than as a fixed structure.

\subsection{Markov Chain Approximation} \label{sec:MarkovApproximation}
We approximate the game dynamics as a Markov chain. The ability to aggregate the agent-based, and by extension, game theoretic models into a Markov chain approximation, is discussed thoroughly in \cite{Banisch2015}. Here, we show how a simplified APT3 (``notional'' in what follows) attack represented as a GPLADD game can be treated as a Markov chain.

Depending on the assumptions about the time-to-success distributions and the attacker and defender strategies, the Markov representation can be exact or approximate. The Markov chain representation of the game dynamics allows a simpler and often practically useful and  sufficient treatment of certain aspects of the defender strategy optimization.

While the definition of a stochastic process state can often be expanded to make the dynamics Markovian, it is likely that not all attacker-defender games can be represented in a practically useful way as Markov chains. For example, a defender may keep a running tally of detections across all attacks by all attackers and drastically change his system when the tally reaches a certain point in a course of a particular attack. By construction, this process will look non-Markovian for the attacker executing the attack when the tally reaches the threshold, because this attacker will not have access to the defender state space representation and the running tally. The possibility of non-Markovian transitions is raised in the context of POMDPs by \cite{Hu2017}. As shown in \cite{Hu2017} this is resolved in POMDPs by introducing the agent belief state, whose updating is Markovian. A similar approach maybe used in the previous example with the defender running a tally of detected attackers, where the attacker may form a belief distribution over possible system states that includes different tally counts.

For the notional attack example in this paper, we create the Markov chain representation of the game between the attacker and the defender by two methods. The first method is used in the Section \ref{sec:sensitivity}. It is based on integrating the time-to-success distributions at each time step to estimate the success probability, estimating the detection probability based on available data or using the available subject-matter expert knowledge, and calculating the probability of the attacker remaining at the same time step by treating the success and detection for each time step as independent events, so that the resulting probability of all three possible events at each time step is 1.

This method then allows evaluating what the defender can improve in his system, by changing his resource allocation across different attack steps. It demonstrates the opportunities and the need for the defender detection strategy optimization.

Our second method is based on directly on the MITRE ATT\&CK Evaluations data for the APT3 attack emulation. It estimates the Markov chain parameters directly from the MITRE ATT\&CK Evaluations data and skips integrating the time-to-success distributions.

The primary difference between the two representations is that the one based on ATT\&CK Evaluations assumes that the probability of staying in the same state between time steps is zero to be consistent with the way the Evaluations are designed.

In the Markov chain approximation of the attack, we assume that the attacker conducts campaigns composed from the individual attacks. Interpreting the Markov chain as an attack campaign allows using the steady state distribution of the Markov to represent the fraction of the time an attacker spends at a particular step.

The attacker runs campaigns either until he achieves the desired success conditions or abandons the attack. The timing of the attacks in the campaign is determined by the attacker strategy.

The methodology for estimating the detection probabilities and Markov chain parameters from the ATT\&CK Evaluations is described in the Section \ref{sec:MITREEvals}. The first method and the notional attack are described in the section \ref{sec:MarkovChainRepresentation} . 

\section{Attack Definition, Attacker and Defender Strategies, Markov Representation, and Results} \label{sec:AttackAndResults}

\subsection{Attack Definition and Markov Representation} \label{sec:example}

We analyze a simplified version of the APT3 techniques as described in the MITRE ATT\&CK Framework. The attack chain starts with a phishing attack, and then pivots through the network with the goal of gaining access to an ICS and taking control of an RTU or a set of RTUs.

The defender network consists of an enterprise network,  an ICS (industrial control system) network, and a DMZ (demilitarized zone subnetwork). The ICS controls an electric power system via RTUs.

This attack goal is to take control of one or more RTUs (remote terminal unit) in an energy distribution system. The attack consists of an  ``initial foothold'' phase aimed at the enterprise network, a pivot phase from the Enterprise network to the ICS,  and the RTU control phase.

\begin{figure}
 \centering
 \includegraphics[keepaspectratio=true, scale=0.40]{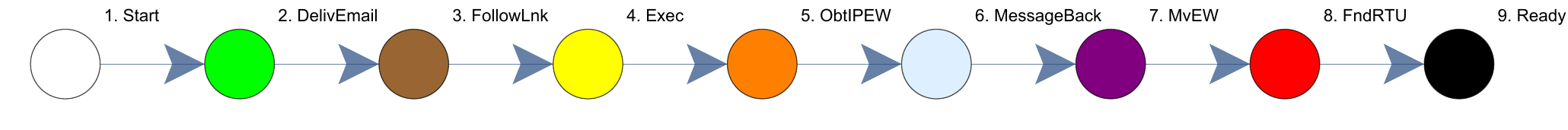}
 \caption{Notional APT3 attack steps.}
 \label{fig:attackSteps}
\end{figure}

In the initial foothold phase, the attacker aims to establish presence on a computer inside of the enterprise network with the goal of discovering an opportunity to pivot to a workstation on the ICS network. The attack steps are visually represented in the Figure \ref{fig:attackSteps}.

The attack begins with the attacker sending a phishing email to induce potential victim to follow a link.

We define a GPLADD game by describing the attack graph, and the attacker and defender strategies. The attack graph is represented as a set of success conditions:

\begin{equation}
\psi= \{c_1, c_2, c_3, c_4, c_5, c_6, c_7, c_8, c_9\},
\end{equation}
where the intermediate attack success conditions, also represented in the Figure \ref{fig:attackSteps}, are as follows, with a short name in () indicating the attack step abbreviated description:

\begin{itemize}
\item $c_1$ - Attack Start (Start).
\item $c_2$ - Email delivered to user (Email).
\item $c_3$ - User follows a link to a malicious website, designated as badguy.com (Link).
\item $c_4$ - The malicious code (trojan) from the website executes on the user computer (Exec).
\item $c_5$ - The trojan obtains ``address'' of an engineering workstation. This is a waiting stage, the user of the infected machine either has access or doesn't have access to the engineering workstation (IPEW).
\item $c_6$ - After engineering workstation connection is established, the trojan sends a message back to the command and control center (Msg).
\item $c_7$ - Once the user connects to the engineering workstation, the trojan moves to the workstation (MvEW).
\item $c_8$ -  The trojan finds RTUs and targets them (RTU).
\item $c_9$ -  When RTUs are found, the trojan is in attack ``Ready'' state. The ``Ready'' state is achieved if at least one RTU is found (Ready).

\end{itemize}

This readiness in the ``Ready'' state denotes the fact that the attacker has successfully completed the infiltration to their desired network location, but before the attacker has executed their final objective (e.g. exfiltration of data or changes to control systems). The fraction of the time when the attacker is ``ready'' to execute on their objectives is represented by the steady state fraction of time the attack spends in the ``Ready'' state.

An attack succeeds if all the terminal conditions are satisfied. For this attack, the attack succeeds when the condition $c_9$ is satisfied.

As explained before, this attack graph does not uniquely determines the attack pathways. The attack can follow a large number of pathways even for this simple single-chain attack graph shown here, depending on the attacker and defender actions. These actions are determined by the attacker and defender strategies, described below.

\subsection{Attacker strategy}
The attacker strategy is composed of active components that are relevant to the interaction with the defender systems and in preparatory components that are relevant as a reflection of the attacker background and other capabilities available for the attack. We treat these background and prepared capabilities as static during the attack and use only information on active components of the strategy in the game analysis.

Additionally, we represent the attacker strategy as two components operating at different time scales: 1) the attack campaign that consists of a number of sequentially executed attacks; and 2) single attack execution strategy.

The purpose of introducing an attack campaign is two-fold: to represent attackers who may attempt multiple attempts at getting into a particular system; and 2) to enable representation of the attack difficulty via the analysis of the corresponding stochastic process and to allow certain parameters of such process, as steady state distribution, expected time to success, etc. to be interpreted as attack metrics. As a part of the attack campaign, the attacker sends emails at certain intervals, determined by their strategy. The emails spread over time to decrease the chance of being noticed by the IDS system of the target organization. We assume that the attacker sends just one email at any given time to avoid detection.

We consider here an ``eager'' attacker strategy, defined in \cite{Outkin2019} as the attacker starting at the next attack step as soon as all the preconditions are satisfied.

\subsection{Defender Strategy}
The defender runs an IDS (Intrusion Detection System) that attempts detecting malicious activity. The specific tools and techniques used by the defender affect the probability of detection at different stages of the attack. These probabilities are also affected by the structure of defender networks and the attacker capabilities. Changes to the structure and operation of the Enterprise and the ICS networks, such as segmenting the network, can be deployed as mitigations to decrease the attack effectiveness or impacts.

If detection occurs, the defender can roll back some or all of the attacker success. In this paper, we assume that detection of the attack at any stage terminates and completely disrupts that particular attack, and as the result the attacker moves back to the ``Start'' step of the attack. Relaxing this assumption and allowing the attacker to persist at previous steps of the attack will not change the analysis methodology but requires a more extensive data gathering effort.

\begin{figure}
 \centering
 \includegraphics[keepaspectratio=true, scale=0.4]{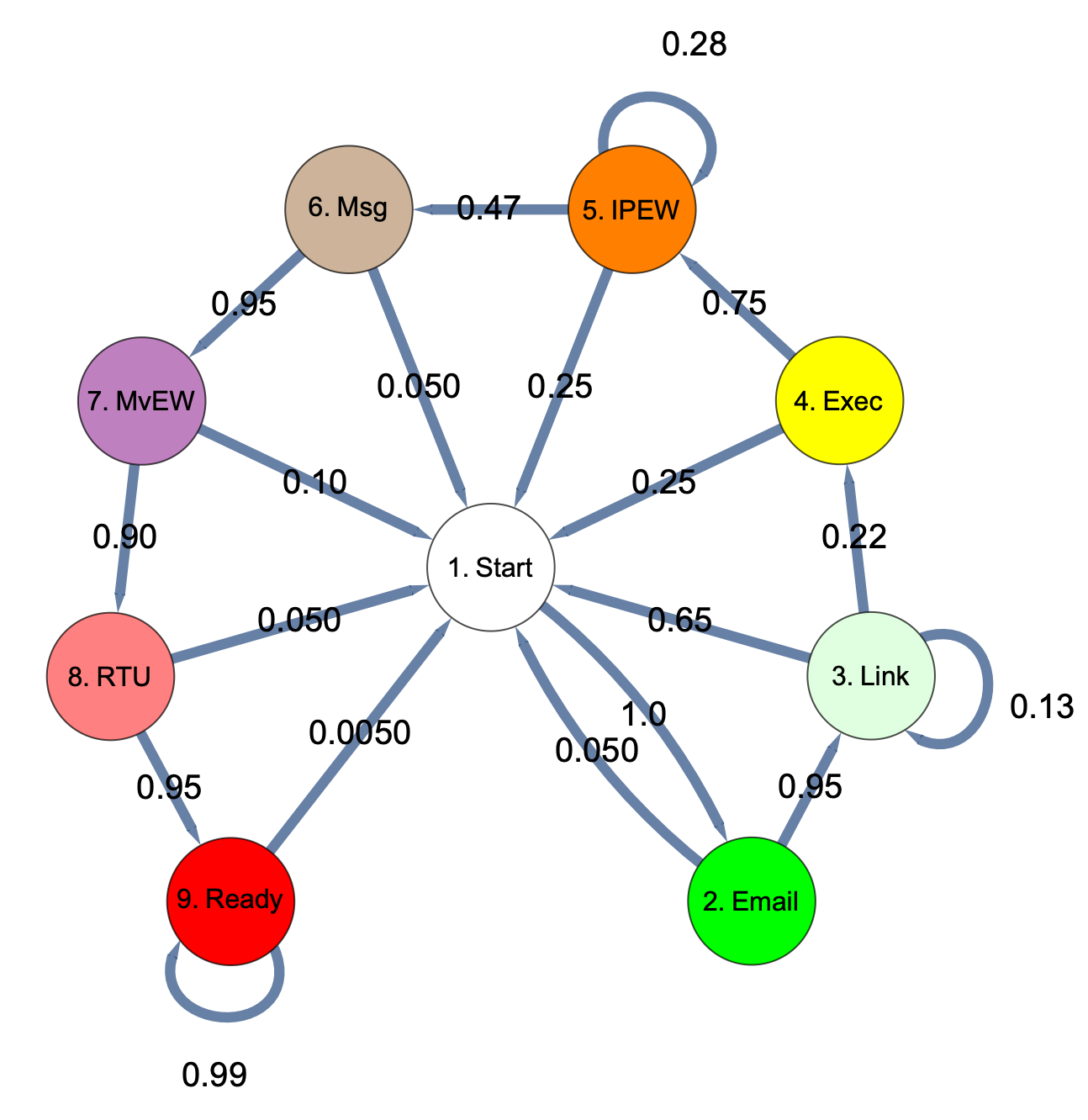}
 \caption{A Markov chain representation of notional attack.}
 \label{fig:Ph1MarkovGraph0_01}
\end{figure}

If the detection does not occur, there are two other possible outcomes: the attacker may succeed or may stay at the same step. Specifying all such probabilities under the Markov assumption that only the current state affects the probability distribution of the future states gives rise to a Markov transitions matrix. This matrix can be represented as a graph showing the transition probabilities between the states. We will call this graph a (Markov) Transition Diagram. An example Markov Transition Diagram, along with the transition probabilities, is displayed in \ref{fig:Ph1MarkovGraph0_01}.

The structure of the Markov chain is the result of the event probabilities in the underlying GPLADD model and of the attacker and defender strategies and capabilities. Its calculation is described in the following section \ref{sec:MarkovChainRepresentation}.

\subsection{Markov Chain Representation} \label{sec:MarkovChainRepresentation} 

This is a description of the first method for inferring the attack success parameters:

\begin{itemize}
\item We introduce a fixed time step duration for all attack steps, $\delta t$. This duration does not need to be the same for different attack steps. It is assumed to be the same for all attack steps for simplicity.

\item We introduce a set of detection probabilities associated with each step in the attack graph. These probabilities are dependent on the time step duration. We also refer to them as time-to-detection probabilities to reflect that the longer the attacker spends on a particular step, the larger the detection probability will be.
    
\item We interpret detections as giving the defender the ability to disrupt the attack and to send the attacker to the very first step or to one of the preceding steps in the attack.

\item Attacker success for each step means moving to the next step and avoiding detection.

\item The attacker may also stay at the same step of the attack in multiple time steps, if they did not succeed in executing the methods required to getting to the next attack step and at the same time has not been detected.

\item Therefore, each step has success (moving to the next step and not getting detected),  failure (detection), and stay-in-place probabilities. The stay-in-place probability is not addressed in the MITRE ATT\&CK Evaluations in this paper.

\end{itemize}

The Markov chain representation is then inferred from the time-to-success distributions and attacker and defender strategies.

The raw probability of success for each attack step is calculated by integrating the corresponding time-to-success distribution $f(t)$ for each step starting at zero and with the corresponding time step duration as the upper bound. The effective probability of success for an attack step is calculated as the raw probability of success times the probability that detection did not occur, under assumption that success and detection are independent random variables.

The time-to-success distribution parameters are based on the SME Evaluations and published research.  For example, according to \cite{williams2018}, about 30\% of phishing emails are opened by their targets. The \cite{williams2018} also describes certain attack success parameters dependence on the attack technique used. This and other information have been incorporated into the attack representation.

In this method, the probability of detection is based on SME information and in some cases on the parameters inferred from published research.

There is also a probability that the attacker will not be detected and will not be able to complete an attack step. This is calculated as one minus the detection and effective success probabilities for the corresponding time step.

This and the approach described above based on the ATT\&CK Evaluations data allow creating a Markov chain representation with the states of the Markov chain based on nodes in $\psi$.

A somewhat different method for creating a Markov chain representation for an attacker defender interaction can be found in \cite{bishop2019Phd}.

We have created a Markov transition matrix based on this method and populated it with transitions probabilities with the $\delta t = 1hr$. This transition matrix is shown in the Table \ref{table:LitSMETransMatrix}.

\begin{table}[ht]
\caption{The Markov process transition matrix showing the transition probability for each attack step. The transition probabilities were inferred from the available literature and/or based on subject matter expert opinions. The time step for which the probabilities were evaluated is 1 hour.}

\centering
\begin{tabular}{l l l l l l l l l l}
  \hline\hline

  Step &      &     &      &      &      &      &      &      &      \\
  \hline
            &1     & 2   & 3    & 4    & 5    & 6    & 7    & 8    & 9    \\
  1         & 0.0  & 1.0 & 0.0  & 0.0  & 0.0  & 0.0  & 0.0  & 0.0  & 0.0  \\
  2         & 0.05 & 0.0 & 0.95 & 0.0  & 0.0  & 0.0  & 0.0  & 0.0  & 0.0  \\
  3         & 0.65 & 0.0 & 0.13 & 0.22 & 0.0  & 0.0  & 0.0  & 0.0  & 0.0  \\
  4         & 0.25 & 0.0 & 0.0  & 0.0  & 0.75 & 0.0  & 0.0  & 0.0  & 0.0  \\
  5         & 0.25 & 0.0 & 0.0  & 0.0  & 0.28 & 0.47 & 0.0  & 0.0  & 0.0  \\
  6         & 0.05 & 0.0 & 0.0  & 0.0  & 0.0  & 0.0  & 0.95 & 0.0  & 0.0  \\
  7         & 0.1  & 0.0 & 0.0  & 0.0  & 0.0  & 0.0  & 0.0  & 0.9  & 0.0  \\
  8         & 0.5  & 0.0 & 0.0  & 0.0  & 0.0  & 0.0  & 0.0  & 0.18 & 0.32 \\
  9         & 0.05 & 0.0 & 0.0  & 0.0  & 0.0  & 0.0  & 0.0  & 0.0  & 0.95 \\

  \hline 
\end{tabular}
\label{table:LitSMETransMatrix}
\end{table}

At this point, they should be treated as rough approximations for three reasons: 1) they are based on the available data and SME opinion about such probabilities in general and as such can only reflect the averages describing the current and recent state of practice in cyber security; 2) they vary significantly between different defender capabilities as shown in the Section \ref{sec:detectionsInference}; 3) they can change with improvements in defensive or offensive capabilities. Certain parameters can also be informed in the future by cyber and emulation-based experimentation and refined with uncertainty quantification.

\subsection{Attack Metrics Sensitivity to Defender Resource  Allocation and Opportunities for the Defender Strategy Optimization} \label{sec:sensitivity}
Our primary goal on this stage of the analysis is to illustrate the ability to use the Markov chain representation of GPLADD games to calculate the attack success metrics and the defender ability to allocate their resources to degrade the attack success metrics.

We show that the defender effort at different parts of his system has substantially different effect on the attack success metrics, on the transitions and the structure of the Markov graph representing the attack, and on the overall attack difficulty. We show that the defender can change the attack transition graph by changing his strategy.

This analysis can assist in success metrics quantification, representing the attack dynamics, attack ranking, identification of improvements to defender system, ranking the improvements, improving defender responses and strategies, system hardening, and mitigations.

\begin{figure}
 \centering
 \includegraphics[keepaspectratio=true, scale=0.250]{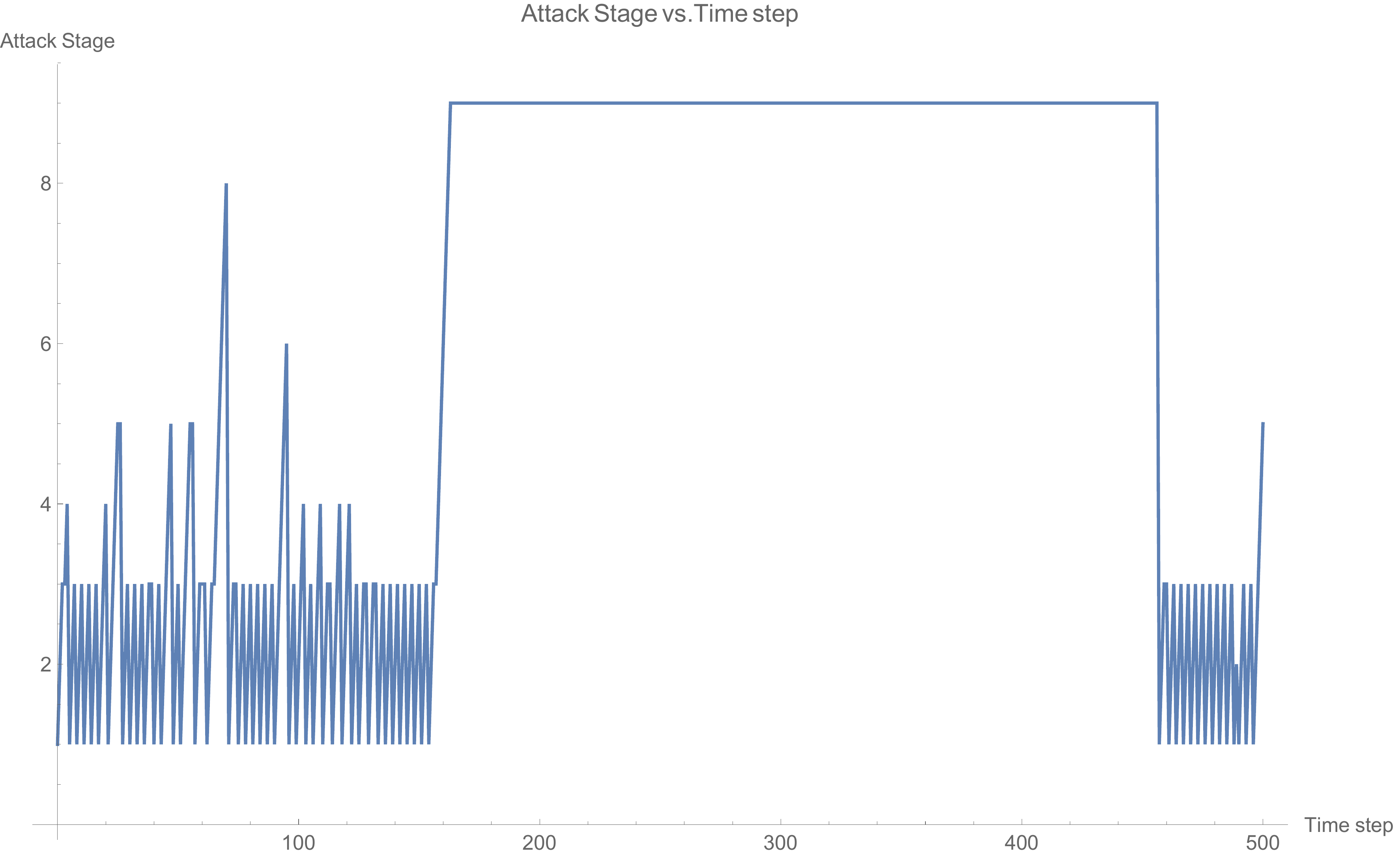}
 \caption{A simulation of attack state over time.}
 \label{fig:ReadyOverTime_p9}
\end{figure}

We start with representation of the attack dynamics over time. Figure \ref{fig:ReadyOverTime_p9} shows an example of simulated time dynamics of an attack, where the attack succeeded sometime before the time step 200, and then was subsequently detected sometime after time step 400, thus allowing the attacker at least 200 time steps to execute either a single, or a set of coordinated attacks. Unlike the fraction of the time the attack spends in the ``Ready'' state, which is based on the steady state of the Markov process, this plot illustrates a single realization of the attack, that may differ between different realizations. In the realization shown here, the attacker spends a lot of time at the steps 1 - 3 of the attack, and only occasionally reaches the steps 4, 5, 6, and 8, only to be quickly dislodged from there. When the attacker finally reaches the step 9 ``Ready'', he stays there for a while, because of the low detection probability in that state.

\begin{figure}
 \centering
 \includegraphics[keepaspectratio=true, scale=0.450]{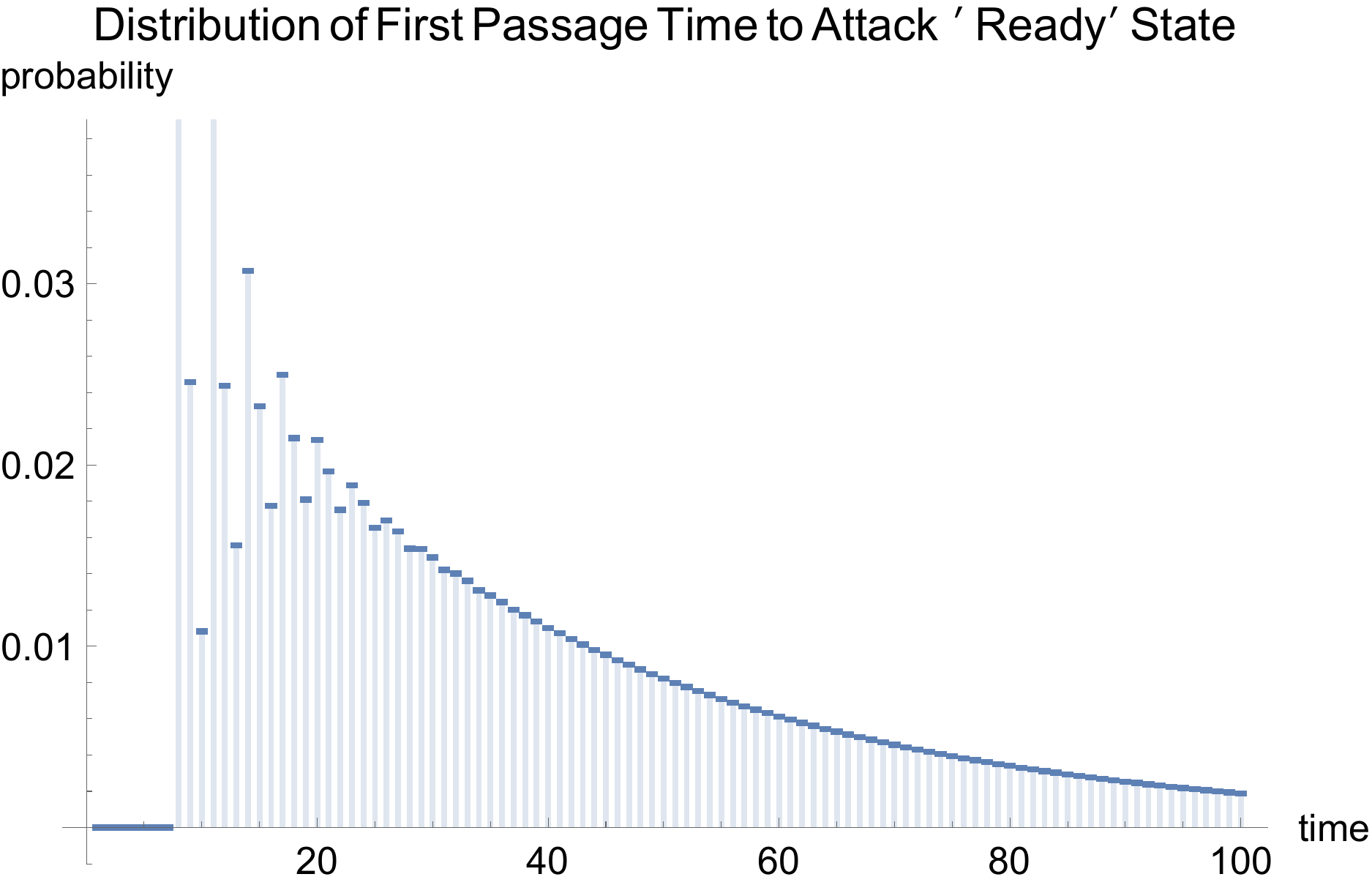}
 \caption{The distribution of the time of the first passage to the ``Ready'' state.}
 \label{fig:TimeToReady}
\end{figure}

Figure \ref{fig:TimeToReady} shows the distribution of the attack first passage time to the ``Ready'' state. The distribution of the first passage time to the ``Ready'' or other states reflects the attack expected progression speed and improve the defender policies.

The time shown on the X-axis is effectively the number of steps in all the attacks done as a part of the attack campaign until success. If a successful attack campaign for a linear attack graph, such as our notional attack, has more steps than the attack graph, it means that one or more of the attacks have failed, and the attacker had to restart the attacks.

Given enough time and an opportunity to conduct the attack campaign even when a set of attacks in the campaign have been disrupted, the attacker eventually succeeds with high probability in this example.

On the other hand, the probability of the attack succeeding undetected (in the case of the notional attack this means succeeding in 9 steps) is small, perhaps in the neighborhood of 4\%.

\begin{figure}
 \centering
 \includegraphics[keepaspectratio=true, scale=0.25]{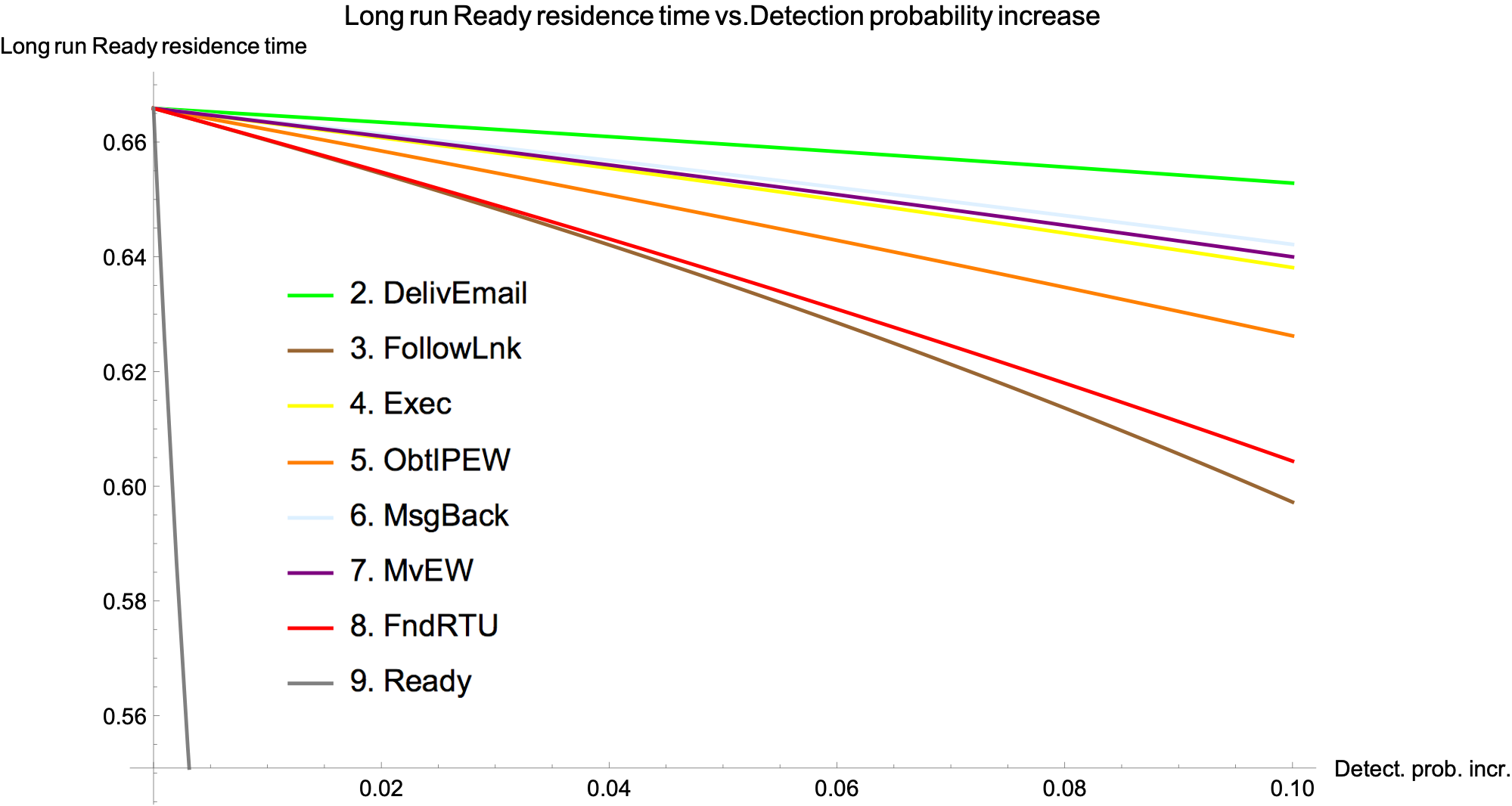}
 \caption{Declines in the attack time spent in the ``Ready'' state vs. investments corresponding to detection at specific stages of an attack.}
 \label{fig:LRReadyVSpdAll_Incr}
\end{figure}

  Figure \ref{fig:LRReadyVSpdAll_Incr} shows how the fraction of time the attack spends in the ``Ready'' state is affected by the improved detection at specific stages of an attack. One clear implication is that from the cost-benefit point of view, it is best to detect the attack at the last, ``Ready'', state. The intuition for this result is very simple: given that the attack spends most time in the ``Ready'' state and little time in any other state, the chances of detecting attack in other states are low, given that most of the time there is nothing to detect there. However, this may not be optimal from the risk-minimization point of view, because if detection at the ``Ready'' state fails, then there are no other options available for the defender to stop the attack. The ability to quantify the effects of improvements in detection corresponding to different steps in the attack, allows the defender to improve their resource allocation and can be used as an input for the defender strategy optimization.
  
We also observe here that making a single step in the attack difficult generally does not have a strong effect on the fraction of the time the Markov chain steady state distribution assigns to the ``Ready'' state, because once that difficult step is passed, there is nothing else that stops the attack from succeeding and we conjecture that changing the defender system so that there are more ways in which the attack may fail or be pushed back to one of the earlier stages may be an effective heuristic for defender optimization.

\subsection{Attack Metrics Comparisons for Different Defender Capabilities} \label{sec:MITREEvals}
MITRE ATT\&CK\textsuperscript{\textregistered} is an industry-standard framework that captures adversary behaviors as they have been observed, and publicly reported, in real world attacks and intrusion campaigns. This framework, and the underlying cyber threat intelligence, informed the MITRE ATT\&CK\textsuperscript{\textregistered} Evaluations (Evaluations). The Evaluations team built a test plan that emulated the cyber intrusion steps demonstrated by a known threat actor, APT3 or Gothic Panda. The emulated scenario was then tested against Endpoint Detection and Response (EDR) products of companies interested in demonstrating their product’s detection capabilities. The released results represent a set of data that can be leveraged to estimate attacker success in similar scenarios or in scenarios that use techniques tested during the Evaluations.

\subsubsection{Markov Process Parameter Inference from MITRE ATT\&CK Evaluations} \label{sec:detectionsInference}
Many steps in the previous GPLADD scenario could be directly mapped to techniques tested in the Evaluations. In this new work, we leveraged Evaluations data to approximate probabilities of detection to inform the Markov chain simulations of GPLADD models. To calculate these probabilities, each step of the scenario in our prior work was mapped to a step in the Evaluations test plan, as outlined in Table \ref{table:defenderCapabilities}. In the Evaluations, each step of the test plan was made up of sub-steps that tested detection for one ATT\&CK Technique. Because our scenario steps more closely aligned with the test plan steps, rather than the sub-steps, we calculated probability of detection for a step as the ceiling of the probabilities of its sub-steps. Probabilities of detection for a sub-step were considered to be the number of detections (detailed below) by unique vendors divided by the total number of vendors (twelve) who participated in the Evaluations.

In the real world, detection depends on the tools and capabilities of the defenders. We tried to capture that dependency by using the “detection categories” defined in the Evaluations and mapping them to a level of defender. The mapping was based on the level of ambiguity that human defenders would be required to deal with. The three ATT\&CK Evaluations categories used in defender capabilities were Indicator of Compromise (IOC), Specific Alert, and General Alert. IOCs provide the highest level of confidence that a system artifact is malicious, as it has been previously identified and labeled as such. Accordingly, the capability to recognize a threat using IOCs requires the lowest investment in defender skills and tools. Specific Alerts signal high confidence that activity or artifacts are malicious, but they may still require investigation by a more skilled defensive team. General Alerts are the most ambiguous of the alert categories, and thus require the most defender expertise when determining malicious activity or artifacts. These defender parameters are summarized in the Table \ref{sec:detectionsInference}.

\begin{table}[ht]
\caption{Detection and Defender Capabilities}
\centering
\begin{tabular}{l l l}
  \hline\hline                  
  Defender Name & Level of Ambiguity & Detection Categories \\ [0.5ex]
  \hline
  Blue 0        & None               & IOC                  \\
  Blue 1        & Medium             & IOC, \\
                &                    & Specific Alerts \\
  Blue 2        & Most               & IOC, Specific Alerts, \\
                &                    & General Alerts \\ [1ex]
  \hline 

\end{tabular}
\label{table:defenderCapabilities}
\end{table}

Each level of defender capability increases the ability to deal with ambiguity, as outlined by the table. Probabilities of detection were calculated, as detailed above, for each of the three detection categories. For a particular level of defender, their overall probability of detection was determined by the ceiling of the probabilities of detection for their relevant detection categories.

We have further considered the two versions of the APT3 attack: 1 and 2 as outlined in the table. Taking some steps from each variation of APT 3 as emulated by the Evaluations team allowed us to create two unique scenarios that mapped to the GPLADD (notional attack) index. Additional information on the MITRE ATT\&CK chains specification and representation can be found in \cite{MITRE2020}. 

\begin{table}[ht]
\caption{Attack Chains used in the analysis.}
\centering
\begin{tabular}{l l l}
  \hline\hline                  
  GPLADD Index                &	Att. Chain 1   &	Att. Chain 2 \\
                              & Steps          &        Steps        \\
  [0.5ex]
  Start	                      &                &                     \\	
  Deliver Email		      &                &                     \\
  Follow Link		      &                &                     \\
  Execute Implant/Malware     &	1.A.1          &	11.A.1       \\
  Obtain Password             &	15.B.1         &	5.A.1        \\
  Message Back (C2)           &	1.C.1          &	11.B.1       \\
  Move to Engineering         &	6.C.1          &	16.E.1       \\
  Workstation                 &                &                     \\
  Find RTU                    &	4.A.1          &	13.A.1       \\
  Ready                       &	7.C.1          &	17.C.1       \\
  [1ex]
\hline 
\end{tabular}
\label{table:detProbByStep}
\end{table}

Applying the methodology described above gives us the probabilities of detection associated with each step for each defender type and each attack variant, as represented in the Table \ref{table:detProbByStep}. We will refer to the combination of defender type and attack variant as Evaluations.  

\begin{table}[ht]
\caption{Detection probability by attack step for each evaluation (E). First number after B is the attack variant (1, 2) and second number is the defender type (0, 1, 2).}
\centering
\begin{tabular}{l l l l l l l l l l}
  \hline\hline                  
  E & & & & &Step & & & & \\ [0.5ex]
  \hline
      & 1 & 2 & 3 & 4   & 5   & 6 & 7   & 8   & 9   \\
  B10 & 0.0 & 0.0 & 0.0 & 0.0   & 0.0   & 0.08 & 0.0   & 0.0   & 0.0   \\
  B11 & 0.0 & 0.0 & 0.0 & 0.58 & 0.08 & 0.17 & 0.0   & 0.0   & 0.67 \\
  B12 & 0.0 & 0.0 & 0.0 & 0.58 & 0.08 & 0.17 & 0.17 & 0.25  & 0.67 \\
  B20 & 0.0 & 0.0 & 0.0 & 0.17 & 0.0   & 0.08 & 0.0   & 0.0   & 0.0   \\
  B21 & 0.0 & 0.0 & 0.0 & 0.75  & 0.5   & 0.17 & 0.0   & 0.0   & 0.42 \\
  B22 & 0.0 & 0.0 & 0.0 & 0.75  & 0.5   & 0.17 & 0.08 & 0.42 & 0.42 \\ [1ex]

\hline 
\end{tabular}
\label{table:detProbByStep}
\end{table}

\subsubsection{Defender Performance Analysis}

We compare and contrast here the attack results analysis for the three kinds of defender and two kinds of attacks described in section \ref{sec:detectionsInference} and specified in the Table \ref{table:detProbByStep}, designated as Evaluations B10 $\ldots$ B22.

Our primary inferences from this analysis are three-fold: 1) we conclude that even state of the art capabilities available on the market at the time of the original Evaluations, may not be able to prevent certain APT threats from succeeding; 2) there is significant variation in outcomes between defenders of different capabilities and outcomes; 3) the ability to increase the difficulty of the attack at a single step is insufficient to make the overall attack difficult. Rather, resource allocation must be optimized across the enterprise based on data corresponding to different steps in possible attacks.

\begin{figure}
 \centering
 \includegraphics[keepaspectratio=true, scale=0.55]{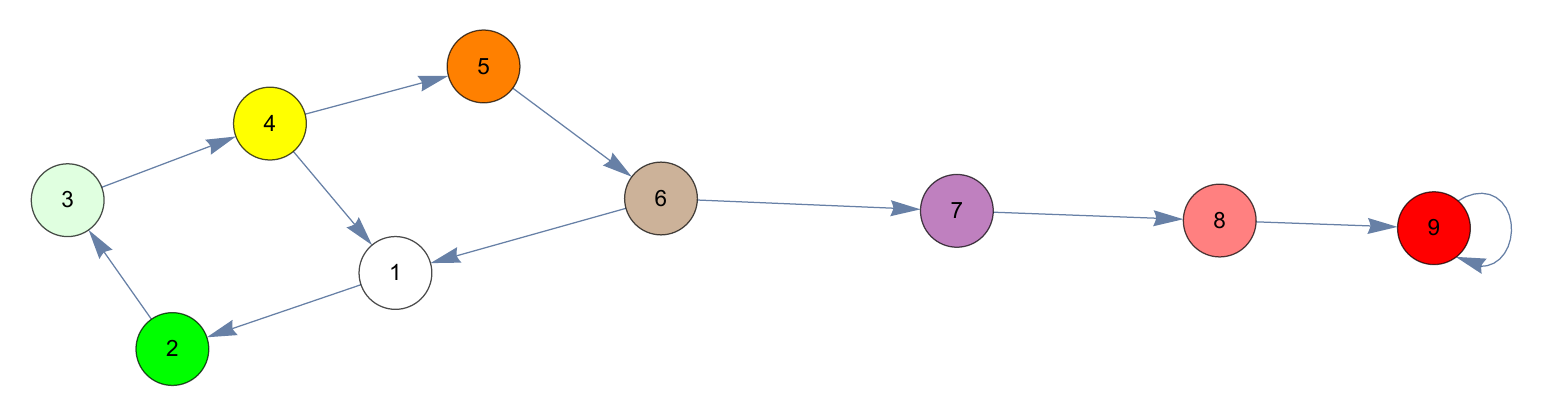}
 \caption{The transition diagram for Defender B0 and Evaluation B20 from Table \ref{table:defenderCapabilities}.}
 \label{fig:TransDiag20}
\end{figure}

\begin{figure}
 \centering
 \includegraphics[keepaspectratio=true, scale=0.55]{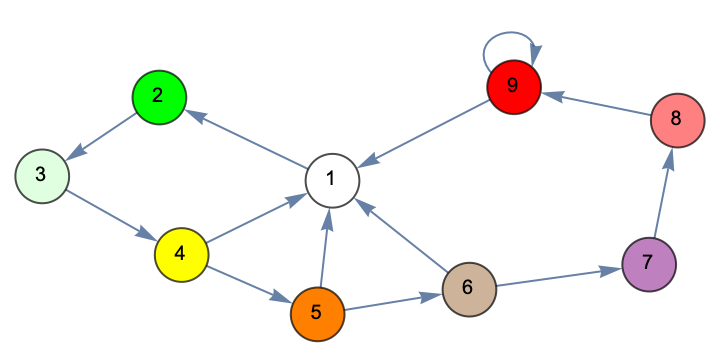}
 \caption{The transition diagram for evaluation B21 for defender B1.}
 \label{fig:TransDiag21}
\end{figure}

\begin{figure}
 \centering
 \includegraphics[keepaspectratio=true, scale=0.75]{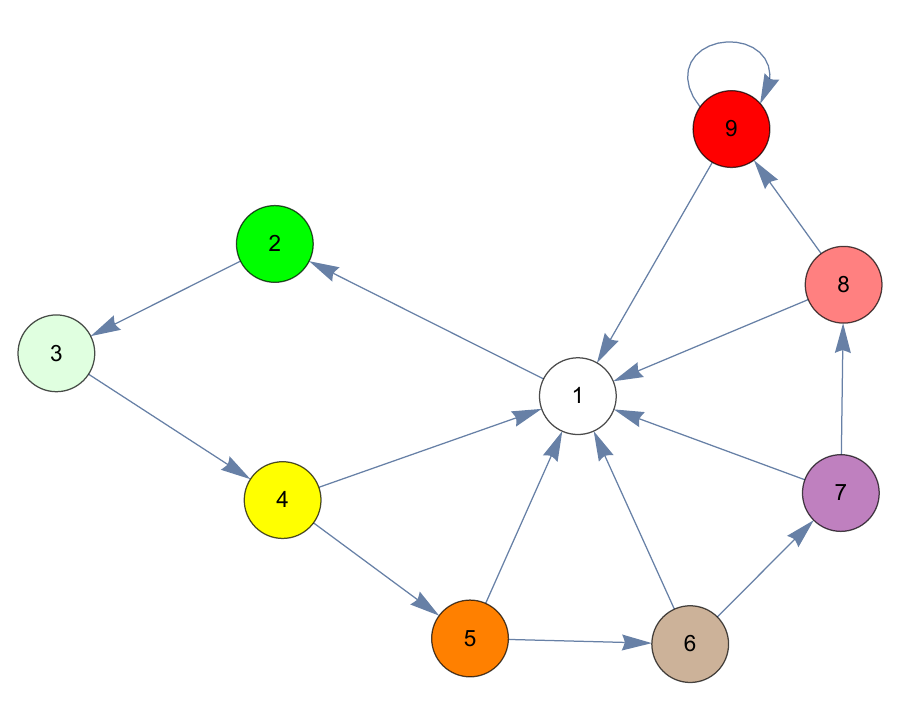}
 \caption{The transition diagram for evaluation B22 for defender B2.}
 \label{fig:TransDiag22}
\end{figure}

We have analyzed the Markov chains corresponding to detection probabilities in the Table \ref{table:detProbByStep}. Figures \ref{fig:TransDiag20} - \ref{fig:TransDiag22} show how the ability to detect the attacker at different steps in the attack changes the structure of the corresponding Transition Diagram. Qualitatively, the results for the attack variants 1 and 2 are very similar, so we describe here the analysis for the attack variant 2.

Figure \ref{fig:TransDiag20} shows the transition diagram for evaluation B20 from Table \ref{table:defenderCapabilities}.

It is clear that in evaluation B20 the defender, B0, is not very capable, because they are able to detect and disrupt the attack only at the earlier attack stages; once the attacker succeeds at the step 6, there is no chance of detecting and arresting the attacker progress anymore.

Figures \ref{fig:TransDiag21} and \ref{fig:TransDiag22} show defenders with improving capabilities. In Evaluation B21 (\ref{fig:TransDiag21}), the defender B1 is better able to detect and respond to threats than defender B0. This is reflected in the fact that the transitions diagram for B21 has more ways that B1 can disrupt the attack than B20.

Defender B2 (Figure \ref{fig:TransDiag22}) has the best capability of all three. They are able to roll back the attacker progress at most steps in the attack.

Understanding of the attack progress can be improved by augmenting it with the quantitative attack success parameters. Given the corresponding Markov chain transition probabilities, we can calculate the steady state of the Markov chain and a measure of the time-to-success distribution. We define the time-to-success distribution as distribution of the first time the attacker reaches the ``Ready'' node.

We first analyze the steady state distributions corresponding to the Evaluations B20 - B22 and see that they paint even more drastic difference between the defenders B0, B1, and B2.

\begin{figure}
 \centering
 \includegraphics[keepaspectratio=true, scale=0.45]{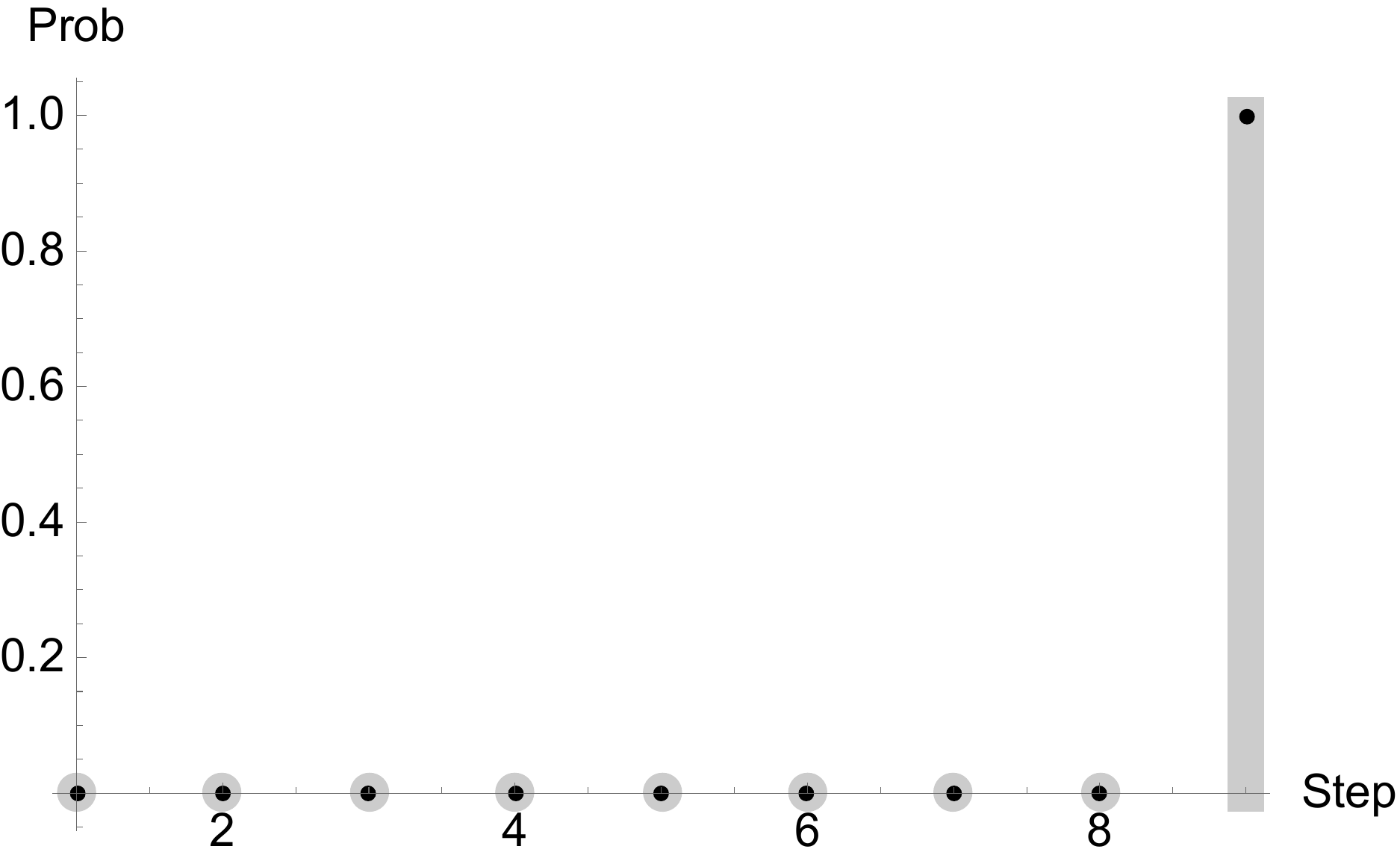}
 \caption{The Markov chain steady state distribution for evaluation B10 and defender B0.}
 \label{fig:SteadyStateDistributionB20}
\end{figure}

B0 effectively has no defense against the attack -  the steady state distribution as shown in Figure \ref{fig:SteadyStateDistributionB20} has all its weight on the ``Ready'' state. This can be interpreted as a reflection of the fact that if the attacker is able to repeat their campaign until step 6 succeeds, the defender has no ability to revert that outcome, unless the defender rebuilds his system from scratch (this option is not considered in this analysis).

This result is consistent with the transition graph shown in the Figure \ref{fig:TransDiag21}.

\begin{figure}
 \centering
 \includegraphics[keepaspectratio=true, scale=0.45]{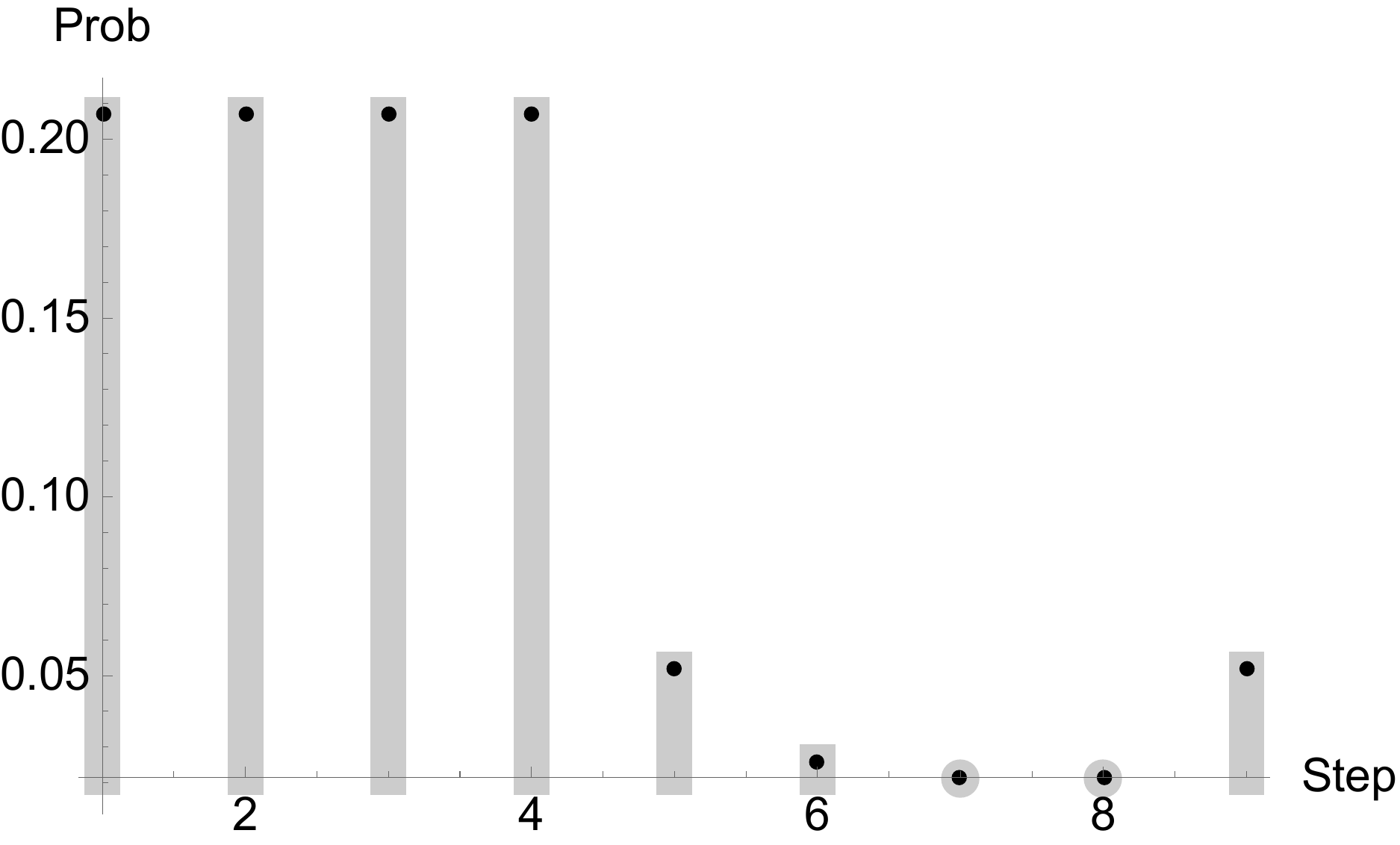}
 \caption{The Markov chain steady state distribution for the evaluation B21 and defender B1.}
 \label{fig:SteadyStateDistributionB21}
\end{figure}

\begin{figure}
 \centering
 \includegraphics[keepaspectratio=true, scale=0.45]{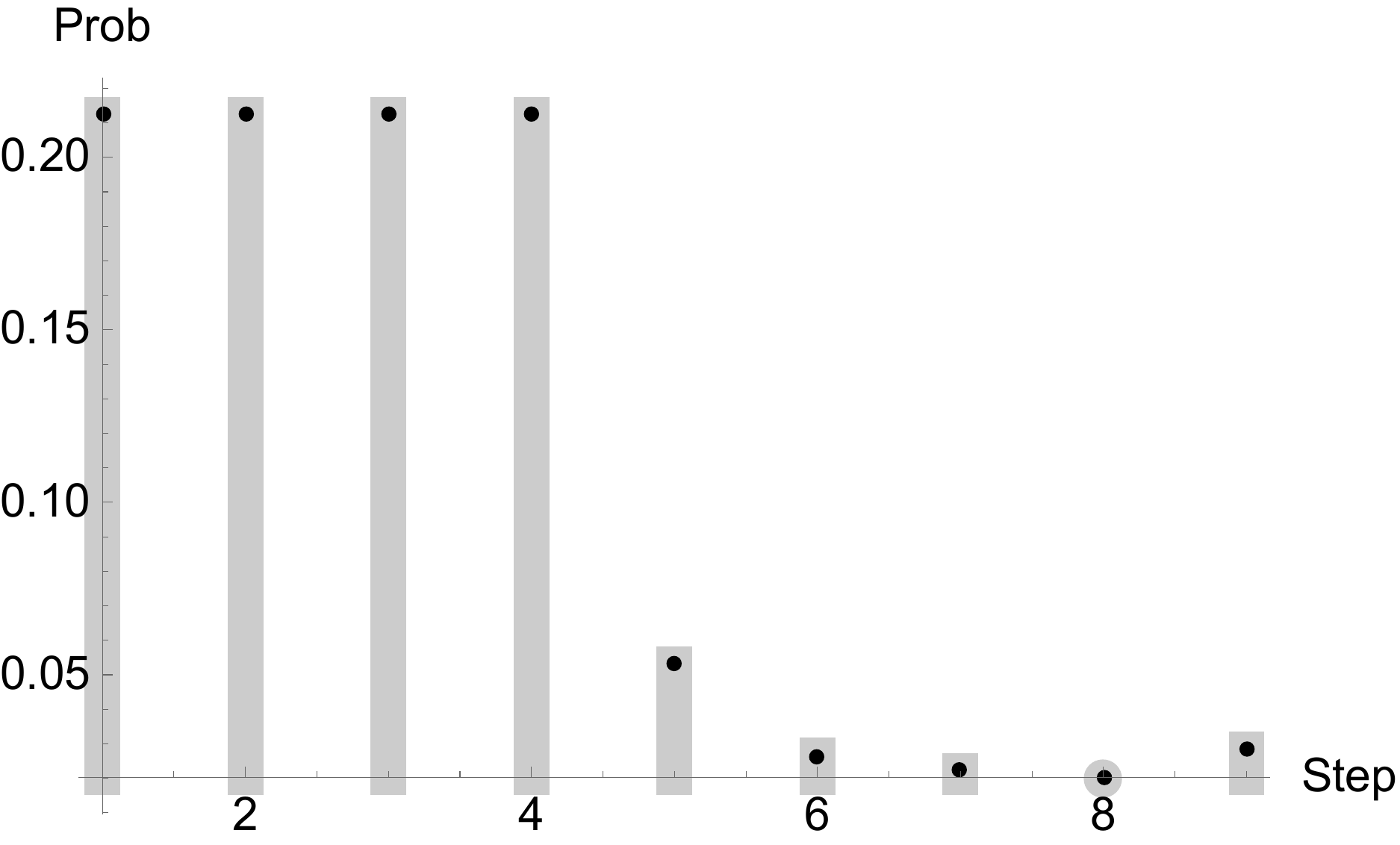}
 \caption{The Markov chain steady state distribution for evaluation B12 and defender B2.}
 \label{fig:SteadyStateDistributionB22}
\end{figure}

When we look at the improved defender capabilities for defenders B1 (\ref{fig:SteadyStateDistributionB21}) and B2 (\ref{fig:SteadyStateDistributionB22}), we see vastly different steady state distributions. In Figure \ref{fig:SteadyStateDistributionB22}, it is the attacker who appears to be a a significant disadvantage.

It is also notable that while Figures \ref{fig:SteadyStateDistributionB21} and \ref{fig:SteadyStateDistributionB22} look similar, the attacker performance at the ``Ready'' state is noticeably worse in Figure \ref{fig:SteadyStateDistributionB22} with about 2\% of the time allocated to the ``Ready'' state as compared to about 5\% of time at the ``Ready'' state in Figure \ref{fig:SteadyStateDistributionB21} . This difference is due to the fact that the defender B1 cannot detect the attacker at stages 7 and 8. It is also a result of weak detection capabilities at stages 7 and 8 (8\% and 42\% respectively) for defender B2. Whether the difference between 5\% vs. 2\% in ``Ready''state is material would be determined by the defender preferences, cost difference between B1 and B2 capabilities, and the nature of the system the defender is protecting. 

\begin{figure}
 \centering
 \includegraphics[keepaspectratio=true, scale=0.45]{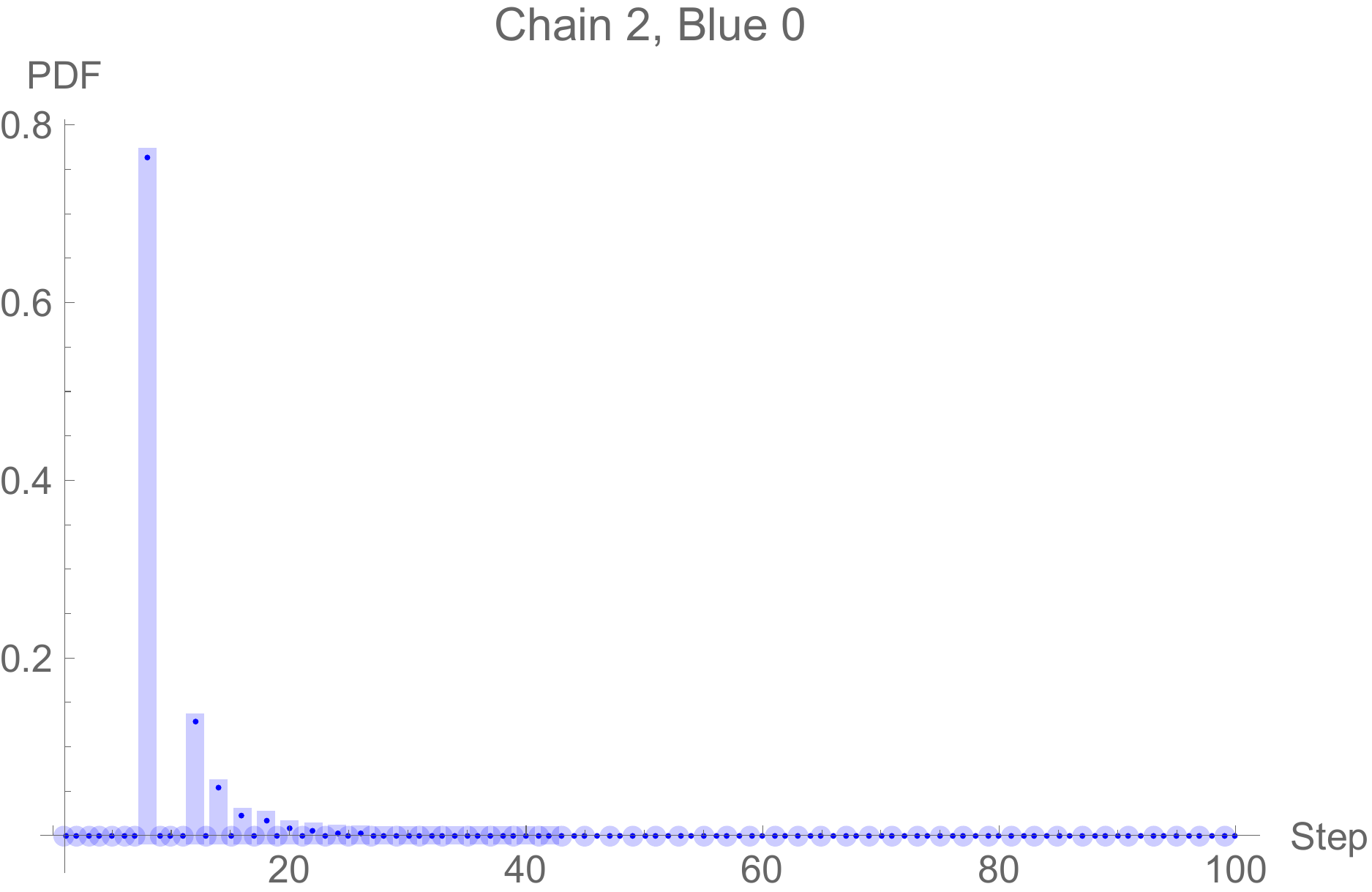}
 \caption{The time to success for evaluation B20 and defender B0.}
 \label{fig:TimeToSuccessB20}
\end{figure}

\begin{figure}
  \centering
  \includegraphics[keepaspectratio=true, scale=0.45]{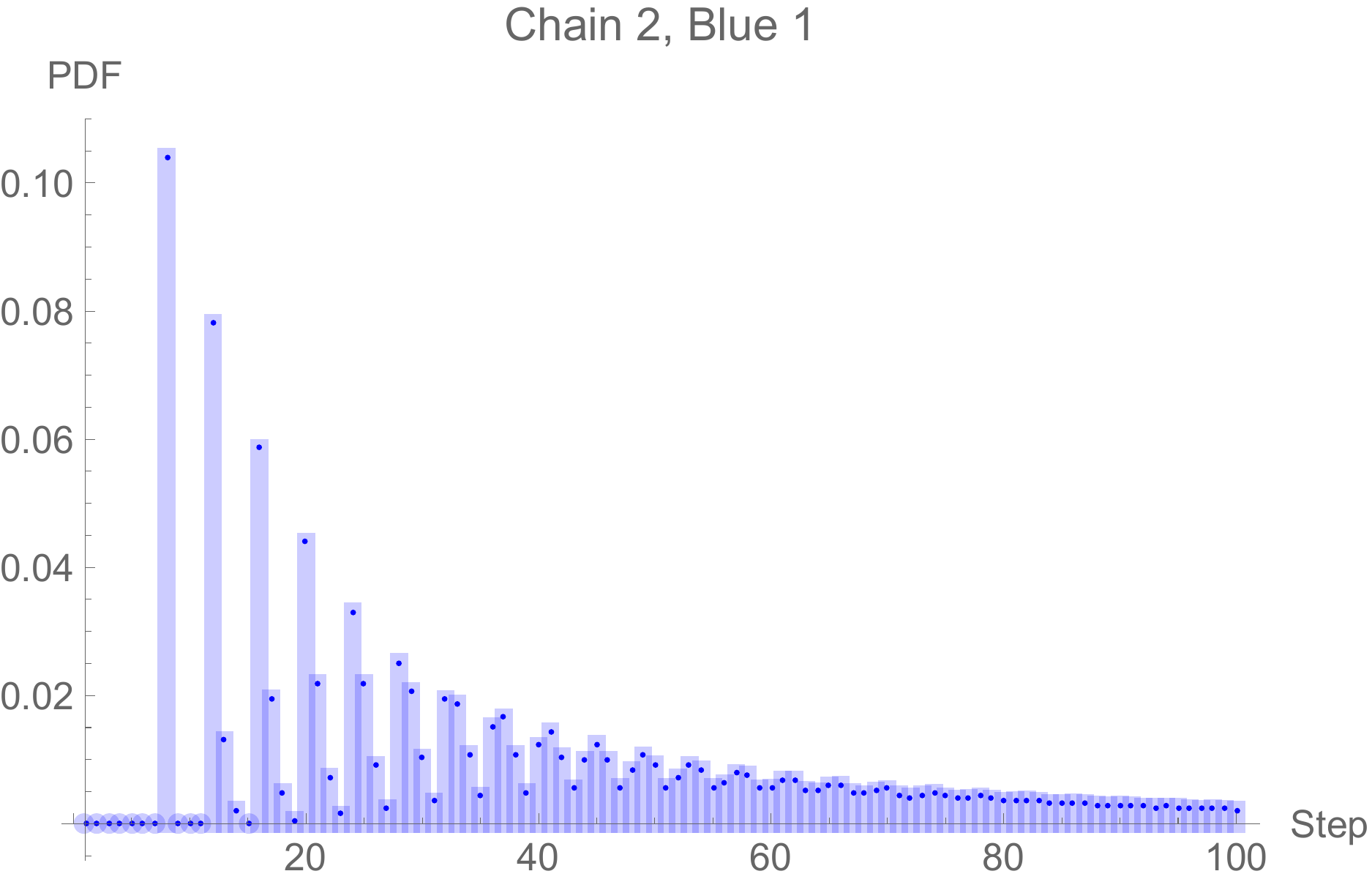}
  \caption{The time to success for the evaluation B21 and defender B1.}
  \label{fig:TimeToSuccessB21}
\end{figure}

\begin{figure}
  \centering
  \includegraphics[keepaspectratio=true, scale=0.45]{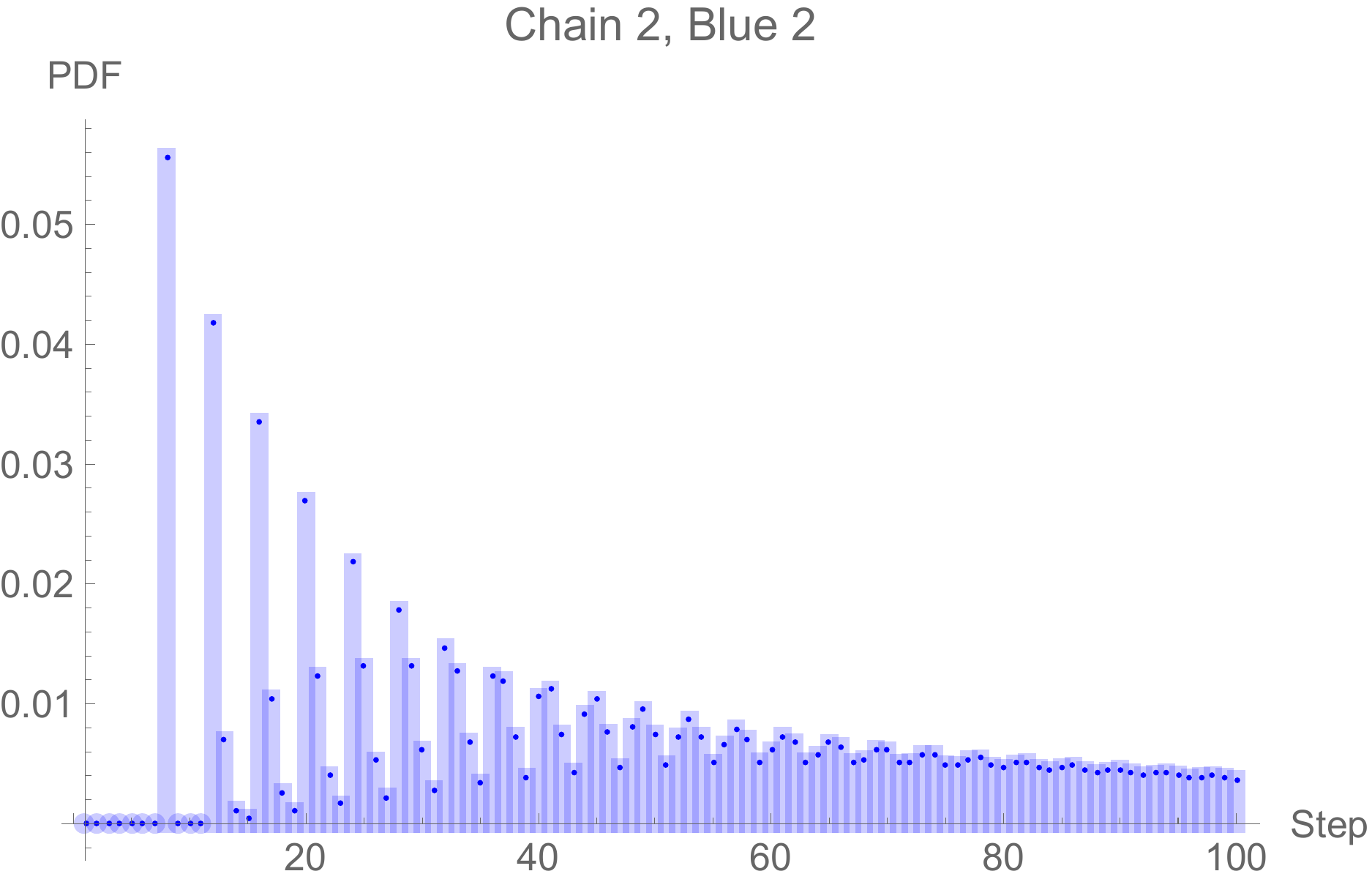}
  \caption{The time to success for the evaluation B22 and defender B2.}
  \label{fig:TimeToSuccessB22}
\end{figure}

When looking at the steady state distributions, it is also important to understand whether they are operationally relevant, because the convergence time to steady state may be long. To investigate that aspect of the notional APT3 attack, we calculated the ``time-to-success''distribution defined as the time of the first passage to the ``Ready'' state. These distributions are shown in Figures \ref{fig:TimeToSuccessB20} - \ref{fig:TimeToSuccessB22}. The times to success shown here are somewhat notional. A key component of the time is to allow artifacts generated by the adversary behaviors to register in the necessary logs that seed defensive information analysis capabilities. The time required for information to percolate from logging to analysis can also vary widely between defenders, but we did not have data to incorporate this variability into our analysis. Despite this limitation of the analysis, we believe these distributions are still informative for defender decision making, because of substantial differences in the shape of the distributions.

In evaluation B20 (Figure \ref{fig:TimeToSuccessB20}), defender B0 is largely unable to stop the attacker progress: about 80\% of the time the attacker succeeds completely unimpeded. 

The improved defender B1 (\ref{fig:TimeToSuccessB21}) and B2 (\ref{fig:TimeToSuccessB22}) capabilities result in substantially better outcomes and time-to-success distributions that are more favorable to the defender.

We see that in the evaluation B21, the attacker succeeds unimpeded with a probability of only about 10\% and in the evaluation B22, this probability further drops to about 6\%.

Given appropriate timing information, the same technique for quantifying the time-to-success distribution for the attack that starts at step 1 can be applied to estimating the attacker progress as a conditional distribution given a delayed detection at a different particular step of the attack. The likely progress the attacker has made between an attacker's action (reflected in logs) and detection by defender (an actionable alert) can then be estimated by applying the attack state at the action/logging time to the corresponding Markov process transition diagram. This generates the distribution of possible states at the detection time and can help direct either automated or manual response.

\begin{figure}
  \centering
  \includegraphics[keepaspectratio=true, scale=0.45]{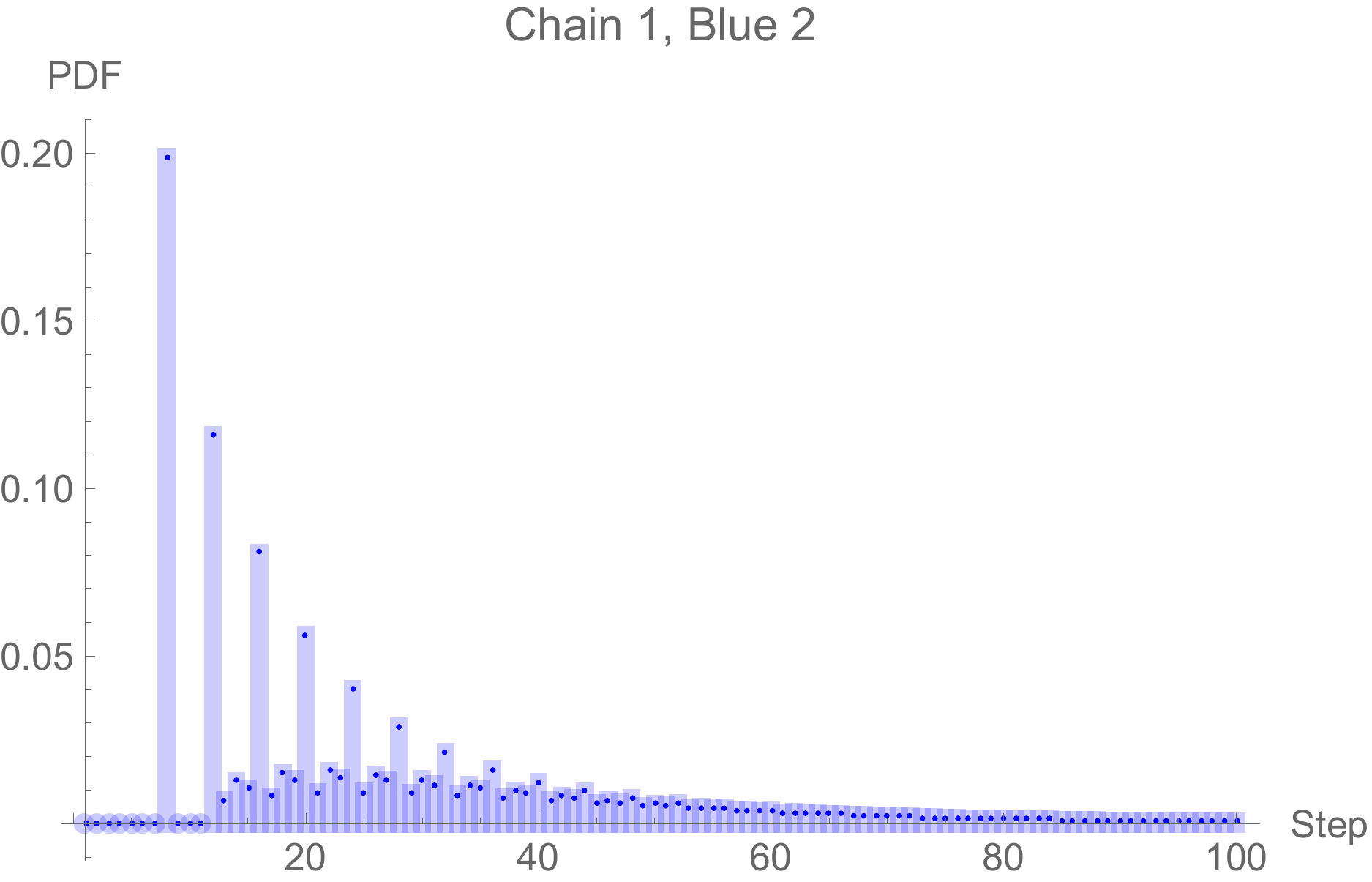}
  \caption{The time to success for the evaluation B12 and defender B2.}
  \label{fig:TimeToSuccessB12}
\end{figure}

We also would like to highlight another, less immediately obvious implication from our analysis that emphasizes the need for defenders to optimize resource allocation between  different attack steps (in real world, this would roughly correlate to investing resources in detection of a variety of TTPs from the ATT\&CK framework). If we compare Evaluations B12 and B22 by looking at the Table \ref{table:detProbByStep}, it is not immediately clear which one would give better outcomes to the defender. The set of steps with positive detection probabilities is the same for both Evaluations, but the detection probabilities themselves are different between B12 and B22. In some cases, B12 has a higher probability of detection vs. B22: in step 7 and especially in step 9 with 67\% detection probability vs. 42\% for evaluation B22.In other steps, e.g. 4 and 8, the detection probability for B12 is lower than for B22. This can be interpreted as B2 allocating more resources to detection on steps 7 and 9 of attack chain 1 and the same defender B2 allocating more resources to steps 4 and 8 for attack chain 2. We can not say a priori which allocation is better, but we can use the time-to-success distribution to investigate the differences between the outcomes defender B2 achieves against attack chains 1 and 2 represented respectively in Evaluations B12 (Figure \ref{fig:TimeToSuccessB22}) and B22 (Figure \ref{fig:TimeToSuccessB12}). We see that even though the evaluation B12 provides the defender with a much better opportunity to detect  the attacker at ``Ready'' (step 9), this alone is insufficient to degrade the attacker success. 

In particular, we see that the attacker succeeds very quickly with probability of about 20\% in evaluation B12 and with probability of only about 6\% in evaluation B22. This means that optimized effort allocation by defenders (B2 in this case)  across different parts of their system may be more important than making a single step substantially more difficult. This result also supports the need to look at the whole defender system in the context of possible attacks and allocate the resources across multiple opportunities for attack disruption, rather than rely on a single robust defense.

We also note that more advanced defenders, B1 and B2, exhibit non-monotonic time-to-success distributions where relatively high values alternate with relatively low values, giving an appearance of a mixture of distributions. We believe this is because the probability of all possible pathways to attack success that have the same length does not need to be a monotonically decreasing function of the pathway length but leave the exploration of this phenomenon to the future work. 

\section{Conclusions}
We developed a model and a method for explicit quantification of attack success parameters and costs based on individual attack steps difficulty and detection probabilities. Our approach started with explicit characterization of the attack success conditions and attacker and defender strategies, and it used Markov chain representation to generate attack success metrics including the ``Ready'' residence time, time-to-success distribution, and steady state distribution.

This allowed evaluating tradeoffs between defender actions, such as detection efforts at different parts of the system and at different steps of the attack.  For a real-world defender, these differences might look like the tradeoff between investing in tools to alert on a larger variety of TTPs or investing in training for the team analyzing the alerts to recognize more subtle signs of adversary actions. These metrics can be used to  allocate resources between detection on different parts of the system and optimize the defender policy against an attack or a set of attacks. These metrics can also inform cyber or emulation-based experiments.

We used the MITRE ATT\&CK\textsuperscript{\textregistered} vendor Evaluations APT3 data to quantify attacks success parameters and difficulty. We showed that performance of different defender models against APT3 notional attack varied greatly and could mean the difference in the attacker always achieving their objectives vs. the attacker spending the most of their time trying to regain a foothold in the system and rarely accomplishing their objectives.

We showed that changes to the defender resource allocation and detection capabilities on different steps of the attack affected the attack success metrics, potentially in a non-intuitive fashion, and compared such metrics for defenders with different capabilities.

We showed that the defender detection expenditures corresponding to different stages of the attack or attacker TTPs have substantially different effects on the attack success parameters and different returns on defender investments. This strengthens the need to look at the whole defender system in the context of possible attacks and allocate the resources across multiple opportunities for attack disruption, rather than rely on a single robust defense.

This approach can be applied to a large set of attacks, including other attacks and attacker capabilities documented in the MITRE ATT\&CK Framework. The resulting attack metrics assessment can be used to develop or improve the defender policy for a given system, or to change the system if necessary to achieve cybersecurity risk reduction or other defender objectives. In addition to MITRE ATT\&CK Evaluations data, this approach can use similar data derived from cyber experimentation in emulated environments and other empirical methods. The analysis in this paper can also allow identifying the parameters whose uncertainty has the greatest effect on attack success metrics and therefore guide the emulation experiments or other data collection efforts.

\bibliographystyle{plain}

\section*{Acknowledgments}

This paper describes objective technical results and analysis. Any subjective views or opinions that might be expressed in the paper do not necessarily represent the views of the U.S. Department of Energy or the United States Government. Sandia National Laboratories is a multimission laboratory managed and operated by National Technology \& Engineering Solutions of Sandia, LLC, a wholly owned subsidiary of Honeywell International Inc., for the U.S. Department of Energy's National Nuclear Security Administration under contract DE-NA0003525. SAND2021-7713.

The authors would like to thank Laura Swiler, Eric Vugrin, Jerry Cruz, Anya Castillo, Zach Benz, Jared Gearhart, Cindy Phillips, and the rest of the SECURE Grand Challenge team for valuable suggestions and comments. We would like to thank Vince Urias for valuable discussions and advice without which this paper may not have been possible. Alexander Outkin would like to thank Chip White, Reid Bishop, and Bob Cole for insightful discussions on Partially Observed Markov Decision Processes and Games, and Rossitza Homan, Mike Livesay, and Dean Jones for valuable discussions on cybersecurity modeling, representation, and visualization, and Jacquilyn Weeks for invaluable writing advice.

\end{document}